\documentclass[12pt,dvipdfmx]{article}

\renewcommand{\thefootnote}{\fnsymbol{footnote}}

\usepackage{amsbsy,amssymb,latexsym,amsfonts,amsmath}
\usepackage{mathrsfs}
\usepackage[pdftex]{graphicx}
\usepackage{bm}
\usepackage{here}

\numberwithin{equation}{section}
\newcommand{\bel}[1]{\begin{equation}\label{#1}}                     
\newcommand{\bal}[1]{\begin{eqnarray}\label{#1}}                     
\newcommand{\be}{\begin{equation}}
\newcommand{\ee}{\end{equation}}
\newcommand{\im}{\mathrm{i}}
\newcommand{\ex}{\mathrm{e}}
\newcommand{\de}{\mathrm{d}}
\newcommand{\dis}{\displaystyle}

\newcommand{\qq}{\qquad}

\newcommand{\dket}{\left<\!\!\!\left<}
\newcommand{\dbr}{\right>\!\!\!\right>}

\begin{document}
%
%
\begin{titlepage}
\begin{flushright}
\normalsize
~~~~
NITEP 1\\
OCU-PHYS 488\\
December, 2018\\
\end{flushright}

\vspace{15pt}

\begin{center}
{\LARGE  Discrete Painlev\'{e} system for the partition function 
 } \\
\vspace{10pt}
{\LARGE of $N_f =2$ $SU(2)$ supersymmetric gauge theory  }\\
\vspace{10pt}
{\LARGE and its double scaling limit}\\
\end{center}

\vspace{23pt}

\begin{center}
{ H. Itoyama$^{a, b,c}$\footnote{e-mail: itoyama@sci.osaka-cu.ac.jp},
T. Oota$^{a,b,c}$\footnote{e-mail: toota@sci.osaka-cu.ac.jp}
  and Katsuya Yano$^b$\footnote{e-mail: yanok@sci.osaka-cu.ac.jp}   }\\

%
\vspace{10pt}
%

$^a$\it Nambu Yoichiro Institute of Theoretical and Experimental Physics (NITEP),\\
Osaka City University\\
\vspace{5pt}

$^b$\it Department of Mathematics and Physics, Graduate School of Science,\\
Osaka City University\\
\vspace{5pt}

$^c$\it Osaka City University Advanced Mathematical Institute (OCAMI)

\vspace{5pt}

3-3-138, Sugimoto, Sumiyoshi-ku, Osaka, 558-8585, Japan \\

\end{center}
%
\vspace{15pt}
\begin{center}
Abstract\\
\end{center}
We continue to study the matrix model of the $N_f =2$ $SU(2)$ case that
represents the irregular conformal block.
What provides us with the Painlev\'{e} system is not the instanton partition function {\it per se}
but rather a finite analog of its Fourier transform that can serve as a generating function.
The system reduces to the extension of the Gross-Witten-Wadia unitary one-matrix model by the logarithmic potential while keeping the planar critical behavior intact.
The double scaling limit to this critical point is a constructive way to study Argyres-Douglas type theory from IR.
We elaborate upon the method of orthogonal polynomials and
its relevance to these problems,
developing it further for the case of a generic
unitary matrix model and that of a special case with the logarithmic potential.


\vfill

\end{titlepage}

\renewcommand{\thefootnote}{\arabic{footnote}}
\setcounter{footnote}{0}

\section{Introduction}

This is a continuation of the paper \cite{IOYanok1} by three of us and is preferably read in sequel.
In the previous letter, we considered the partition function of $N_{f}=2$, $SU(2)$, $\mathcal{N}=2$ supersymmetric gauge theory in four dimensions
by the one-matrix model obtained from a generic irregular conformal block \cite{Gai, MMM09092, GT} as an integral representation \cite{IOYone}. See also \cite{DF,MMS10,IO5} for $N_f = 4$ conformal case.
We derived the discrete Painlev\'{e} system from a set of recursion relations for $R_n$ which is directly related to the finite $N$ partition function. (See eq. \eqref{PartZR} of this paper.)
Here, $N$ is the size of the matrix and, at the same time, the sum of the masses of the two flavors $ -(m_1 +m_2)/g_s$.
Taking the double scaling limit \cite{BK, DS,GM} to the critical point of this system, we succeeded in deriving the Painlev\'{e} II differential equation containing the parameter $M \equiv \alpha_{1+2} +N=  (m_2 -m_1)/g_s$.
See \cite{AGT, IO5, IOYone, IOYanok1} and Appendix \ref{CBandIL} of this paper for detail of these transcriptions or matrix model/4d gauge theory dictionary.
This critical point is the simplest example of the Argyres-Douglas type points \cite{AD, APSW, KY, IOYanok1} well studied by using the Seiberg-Witten curve of cubic or quartic type \cite{SW, HO}.
From the point of view of matrix models per se, this is the extension of the Gross-Witten-Wadia model \cite{GW80,wad1212,wad80} 
by the addition of a logarithmic potential that turns out to keep the planar critical behavior intact.
Here in this paper, we fill in the details of these derivations by elaborating upon the method of orthogonal polynomials \cite{bes79, IZ80,PS90a, PS90b} which we further develop here in the case of a generic unitary matrix model and that of a special case relevant to the problem here where the potential contains a logarithm.

The original gauge theory partition function is an instanton partition function \cite{Nek, NY} of
${\cal N} =2$ $SU(2)$ theory which contains a Coulomb-moduli $=$ filling fraction
parameter.
It is this partition function which is precisely identified with the irregular conformal block by the $2$d-$4$d connection via matrices.
The partition function which is represented as a discrete Painlev\'{e} system is, however, NOT of this kind,
but instead given by a finite $N$ analog of the Fourier transform \cite{IOYanok1}.
This important difference appears to provide a new arena of escaping from instanton vacua
to move to the construction of a more general generating function.
This last point has been discussed in \cite{IOYanok1}, taking the corresponding structure
\cite{GIL12, GIL13, ILT,nag1611, BLMST, GG,LNR1806,AJJRT1607} and the reasoning \cite{MM} at $N_f = 4$ as a hint.
We elaborate upon this in Appendix \ref{CBandIL} of this paper as well.

Most generically, let us consider the following type of the $\beta$-deformed ``matrix model'' depending on two integration contours $C_L$ and $C_R$:
\bel{TCMM}
Z(N_L, N_R) = \mathcal{C}
\left( \prod_{i=1}^{N_L} \int_{C_L} \de w_i \right)
\left( \prod_{j=1}^{N_R} \int_{C_R} \de w_{N_L+j} \right) \Delta^{2\beta}(w) 
\exp\left( \sqrt{\beta} \sum_{I=1}^{N_L+N_R} W(w_I) \right),
\ee
where $\mathcal{C}$ is a normalization constant and $\Delta(w)$ is the Vandermonde determinant:
\be
\Delta(w) = \prod_{1 \leq I<J \leq N_L+N_R} (w_I - w_J).
\ee
Following \cite{MM}, we introduce their generating function
\bel{GTCMM}
\begin{split}
\underline{Z}(N; \mu_L, \mu_R) 
&= \sum_{N_L+N_R=N} \frac{\mu_L^{N_L}}{N_L!} \frac{\mu_R^{N_R}}{N_R!} \, Z(N_L,N_R) \cr
&= \frac{\mathcal{C}}{N!} \int_{C} \de^N w \, \Delta^{2\beta}(w) \, 
\exp\left( \sqrt{\beta}\sum_{I=1}^{N} W(w_I) \right),
\end{split}
\ee
where $C=\mu_L C_L + \mu_R C_R$, i.e., 
\be
\int_C = \mu_L \int_{C_L} + \mu_R \int_{C_R}.
\ee

As we recall from \cite{IOYanok1} and elaborate further in Appendix \ref{CBandIL},
in the case of $\beta=1$ which we consider in the body of this paper, the $N_{f} =2$
matrix model of the above form with $\alpha_{1+2} \in \mathbb{Z}$ in fact reduces to the unitary matrix model with cosine $+$ logarithmic potential in section 3.

In the next section, we consider a generic unitary one-matrix model and
utilize the method of orthogonal polynomials.
In section 3, we study properties of the unitary matrix model with the above potential
at finite $N$ in depth: these are string equations as a set of recursion relations, small $\underline{g}{}_s$ expansion, and the free energy.
In section 4, we consider the planar and
the double scaling limit of the model, which leads us to the derivation of the Painlev\'{e} II equation.
In Appendix \ref{VUN}$\sim$\ref{IEforN2MM}, we elaborate upon several contents in the text, and review a few of the past work of relevance.

 It is instructive to count the number of free parameters, starting from
the $N_{f}=4$ case.
 Letting aside the coupling constant (or the cross ratio), we have
six parameters in the original $0$d-$4$d dictionary \cite{IO5}.
 After sending two of the four mass parameters to infinity, we have four parameters.
 By the transition from the instanton
partition function to the generating function denoted in this paper by $\underline{Z}$,
we get rid of the Coulomb moduli parameter.
 Finally setting $\beta =1$, the number of
free parameters reduces to two.
 These are nothing but the sum and the difference
of the two remaining mass parameters,
 letting aside the QCD like scale parameter
obtained by the dimensional transmutation.

\section{Unitary matrix model}

In this section, we briefly review the unitary matrix model, 
the method of orthogonal polynomials and the string equations to explain our notation.

The partition function of the unitary matrix model is defined by
\bel{PartFn}
Z_{U(N)}:= \frac{1}{\mathrm{vol}(U(N))} \int [\de U]\,
\exp\Bigl( \mathrm{Tr}\, W_{U} (U) \Bigr),
\ee
where $U$ is an $N \times N$ unitary matrix and $W_{U}(U)$ is a potential.
We define a unitary Haar measure $[\de U]$ from the metric
\bel{UNmet}
\de s^2 = \mathrm{Tr}\bigl( \de U^{\dag}\, \de U \bigr) = - \mathrm{Tr}\bigl( U^{-1} \de U \bigr)^2.
\ee
With this normalization of the measure, the volume of the unitary group $U(N)$ is given by  (see Appendix \ref{VUN} for derivation)
\bel{volUN}
\mathrm{vol}(U(N)) = \int [ \de U] = \frac{(2\pi)^{(1/2)N(N+1)}}{G_2(N+1)},
\ee
where $G_2(z)$ is the Barnes function. Explicitly, $G_2(N+1)$ is given by
\be
G_2(N+1) = \prod_{j=1}^{N-1} j! = \prod_{k=1}^{N-1} k^{N-k}.
\ee

If we diagonalize the unitary matrix $U$ as
\be
U = V^{-1} U_D V, \qq
U_D = \mathrm{diag}(z_1, z_2, \dotsm, z_N), \qq
| z_i | = 1,
\ee
we have
\bel{PartFn2}
Z_{U(N)} = \frac{1}{N!} \left( \prod_{i=1}^N \oint \frac{\de z_i}{2\pi \im \, z_i} \right)
\Delta(z) \Delta(z^{-1}) \exp\left(  \sum_{i=1}^N
W_{U}(z_i) \right),
\ee
where
\be
\Delta(z) = \prod_{1 \leq i<j \leq N} (z_i - z_j), \qq
\Delta(z^{-1}) = \prod_{1 \leq i<j \leq N}(z_i^{-1} - z_{j}^{-1} ).
\ee
Let
\bel{measure}
\de \mu(z):= \frac{\de z}{2\pi \im\, z} \exp\Bigl( W_{U}(z) \Bigr).
\ee
Then
\be
Z_{U(N)} = \frac{1}{N!} \int \prod_{i=1}^N \de \mu(z_i)\, \Delta(z) \Delta(z^{-1} ).
\ee
The partition function \eqref{PartFn2} expressed in eigenvalue integrals may be generalized to the form of the
two-contour model \eqref{TCMM}. A natural choice of the two contours $C_L$ and $C_R$ is
to take them as circles of radius $r_L$ and $r_R$ respectively. Suppose, $r_L < r_R$ and
there is no singularity in the region $r_L \leq | w | \leq r_R$. Then the contours can be smoothly deformed
to circles of the same radius, i.e., to the same contour: $C_L = C_R$. Then, for the two-contour unitary matrix 
model, $Z_{U(N)}(N_L, N_R)$ depends only on $N=N_L+N_R$, and the generating function $\underline{Z}{}_{U(N)}$
is essentially $Z_{U(N)}(N,0)$. Because $\underline{Z}{}_{U(N)} = (\mu_L+\mu_R)^N Z_{U(N)}$,
we can set $\mu_L + \mu_R=1$ without loss of generality. Hence $\underline{Z}{}_{U(N)} = Z_{U(N)}$.

\subsection{Orthogonal polynomials}

The unitary matrix model can be solved  \cite{PS90a,PS90b,MP90} by the method of orthogonal polynomials \cite{bes79,IZ80}. For orthogonal polynomials on a unit circle, see, for example, \cite{ger54,ger60, sim1},
and references therein\footnote{The raising and lowering operators for orthogonal polynomials on the unit circle
are considered in \cite{IW0012}.}.
We will use the monic orthogonal polynomials \cite{PS90a,PS90b}\footnote{In \cite{MP90},
orthogonal polynomials of different type have been introduced
to solve the unitary matrix model.}. (See also \cite{sim1}.)

\subsubsection{Definitions and properties}

Let $p_n$ and $\tilde{p}_n$ ($n \geq 0$) be monic polynomials satisfying orthogonality conditions with
respect to the measure \eqref{measure}
\bel{OC}
\int \de \mu(z) p_n(z) \tilde{p}_m(1/z) = h_n \delta_{n,m},
\ee
where
\bel{OP}
p_n(z) = z^n + \sum_{k=0}^{n-1} A_k^{(n)} z^k, \qq
\tilde{p}_n(1/z) = z^{-n} + \sum_{k=0}^{n-1} B_k^{(n)} z^{-k}.
\ee
Let us introduce the moments $\mu_n$ for the measure \eqref{measure} by
\bel{moments}
\mu_n:= \int \de \mu(z) z^n, \qq
(n \in \mathbb{Z} ).
\ee
For later convenience, we define $\mathcal{K}^{(n)}_k$  by
\be
\mathcal{K}^{(n)}_k:= \det \bigl( \mu_{j-i+k} \bigr)_{1 \leq i,j \leq n}, \qq
( n \geq 0, k \in \mathbb{Z}).
\ee
From the definition, the orthogonal polynomials have the following properties:
\bel{pp1}
\int \de \mu(z)\, p_n(z) z^{-k} = 0, \qq
(k=0,1,\dotsm, n-1),
\ee
\bel{pp2}
\int \de \mu(z)\, z^k \tilde{p}_n(1/z) = 0, \qq
(k=0,1,\dotsm, n-1).
\ee
Using these and the monic properties, the orthogonal polynomials are determined as
\be
p_n(z) = \frac{1}{\tau_n} 
\begin{vmatrix}
\mu_0 & \mu_1 & \mu_2 & \dotsm & \mu_n \cr
\mu_{-1} & \mu_0 & \mu_1 & \dotsm & \mu_{n-1} \cr
\vdots & \vdots & \vdots & \ddots & \vdots \cr
\mu_{-n+1} & \mu_{-n+2} & \mu_{-n+3} & \dotsm & \mu_1 \cr
1 & z & z^2 & \dotsm & z^n 
\end{vmatrix},
\ee
\be
\tilde{p}_n(1/z) = \frac{1}{\tau_n} 
\begin{vmatrix}
\mu_0 & \mu_{-1} & \mu_{-2} & \dotsm & \mu_{-n} \cr
\mu_{1} & \mu_0 & \mu_{-1} & \dotsm & \mu_{-n+1} \cr
\vdots & \vdots & \vdots & \ddots & \vdots \cr
\mu_{n-1} & \mu_{n-2} & \mu_{n-3} & \dotsm & \mu_{-1} \cr
1 & z^{-1} & z^{-2} & \dotsm & z^{-n} 
\end{vmatrix},
\ee
where
\be
\tau_n:=\mathcal{K}^{(n)}_0 = \det\bigl( \mu_{j-i} \bigr)_{1 \leq i,j \leq n}.
\ee
(We set $\tau_0 = 1$). We can easily see that these polynomials obey \eqref{pp1} or \eqref{pp2}.
For example,
\be
\int \de \mu(z)\, p_n(z) z^{-k} =
\frac{1}{\tau_n}
\begin{vmatrix}
\mu_0 & \mu_1 & \mu_2 & \dotsm & \mu_n \cr
\mu_{-1} & \mu_0 & \mu_1 & \dotsm & \mu_{n-1} \cr
\vdots & \vdots & \vdots & \ddots & \vdots \cr
\mu_{-n+1} & \mu_{-n+2} & \mu_{-n+3} & \dotsm & \mu_1 \cr
\mu_{-k} & \mu_{1-k} & \mu_{2-k} & \dotsm & \mu_{n-k} 
\end{vmatrix}=0,
\ee
for $0 \leq k \leq n-1$.

The normalization constants $h_n$ defined by \eqref{OC} are given by 
\be
h_n = \frac{\tau_{n+1}}{\tau_n} = \frac{\mathcal{K}^{(n+1)}_0}{\mathcal{K}^{(n)}_0}.
\ee
The constant terms of these polynomials will play important roles.
\be
A_n:=p_n(0) = A^{(n)}_0 =(-1)^{n} \frac{\mathcal{K}^{(n)}_1}{\mathcal{K}^{(n)}_0},
\qq
B_n:=\tilde{p}_n(0) = B^{(n)}_0 = (-1)^n \frac{\mathcal{K}^{(n)}_{-1}}{\mathcal{K}^{(n)}_0}.
\ee
Note that
\be
\frac{h_n}{h_{n-1}} = \frac{\tau_{n+1} \tau_{n-1}}{\tau_{n}^2}, \qq
1 - \frac{h_n}{h_{n-1}} = \frac{\tau_n^2 - \tau_{n+1} \tau_{n-1}}{\tau_n^2}.
\ee
Using an identity
\be
\tau_n^2 - \tau_{n+1} \tau_{n-1} = \bigl( \mathcal{K}^{(n)}_0 \bigr)^2 - \mathcal{K}^{(n+1)}_0 \mathcal{K}^{(n-1)}_0
= \mathcal{K}^{(n)}_1 \mathcal{K}^{(n)}_{-1},
\ee
we can show that
\be
1 - \frac{h_n}{h_{n-1}} = \frac{\mathcal{K}^{(n)}_1 \mathcal{K}^{(n)}_{-1}}{(\mathcal{K}^{(n)}_0)^2}
= A_n B_n.
\ee
Thus we have the following relations:
\be
\frac{h_n}{h_{n-1}} = 1 - A_n B_n.
\ee

\subsubsection{Partition function and the orthogonal polynomials}

Note that
\be
 \prod_{1 \leq i< j \leq N} (z_j - z_i)
=\det \bigl( p_{j-1}(z_i) \bigr)_{1 \leq i,j \leq N}
=\sum_{\sigma \in S_N} (-1)^{\varepsilon(\sigma)} \prod_{k=1}^N p_{\sigma(k)-1}(z_k),
\ee
\be
\prod_{1 \leq i<j \leq N} ( z_j^{-1} - z_i^{-1} )
=
\det \bigl( \tilde{p}_{j-1}(1/z_i) \bigr)_{1 \leq i,j \leq N}
=\sum_{\sigma \in S_N} (-1)^{\varepsilon(\sigma)} \prod_{k=1}^N \tilde{p}_{\sigma(k)-1}(1/z_k).
\ee
Using these relations, the partition function \eqref{PartFn2} is evaluated as
\be
\underline{Z}_{U(N)} = \frac{1}{N!} \int \prod_{i=1}^N \de \mu(z_i) \, \Delta(z) \Delta(z^{-1})
= \prod_{k=0}^{N-1} h_{k}= \prod_{k=0}^{N-1} \frac{\tau_{k+1}}{\tau_k} = \tau_N.
\ee
Also, it can be written as
\bel{uZUN}
\underline{Z}_{U(N)} = h_0^N \prod_{j=1}^{N-1} \bigl(1 - A_j B_j \bigr)^{N-j}.
\ee

\subsubsection{Recursion relations for orthogonal polynomials}

If we expand $z\, p_n(z)$ in the $\{ p_k(z) \}$ basis, 
all of the lower degree polynomials are generated:
\bel{recp1a}
z \, p_n(z) = p_{n+1}(z) + \sum_{k=0}^n C^{(n)}_k \, p_{k}(z),
\ee
where
\be
C^{(n)}_k = (-1)^{n-k} \frac{\mathcal{K}^{(n+1)}_1 \mathcal{K}^{(k)}_{-1} }
{\mathcal{K}^{(n)}_0 \mathcal{K}^{(k+1)}_0 }
=- \frac{h_n}{h_k} A_{n+1} B_k, \qq
(0 \leq k \leq n).
\ee

Similarly, we have for $\tilde{p}_n$
\bel{recp2a}
z^{-1} \tilde{p}_n(1/z) = \tilde{p}_{n+1}(1/z) + \sum_{k=0}^n \widetilde{C}^{(n)}_k \tilde{p}_k(1/z),
\ee
where
\be
\widetilde{C}^{(n)}_k = (-1)^{n-k} 
\frac{\mathcal{K}^{(k)}_1 \mathcal{K}^{(n+1)}_{-1}}{\mathcal{K}^{(k+1)}_0 \mathcal{K}^{(n)}_0}
= - \frac{h_n}{h_k} A_k B_{n+1}, \qq
( 0 \leq k \leq n).
\ee

The above relations \eqref{recp1a} and \eqref{recp2a} can be rewritten as three-term relations:
\be
p_{n+1}(z) = z\, p_n(z) + A_{n+1} \, z^n\, \tilde{p}_n(1/z),
\ee
\be
\tilde{p}_{n+1}(1/z) = z^{-1}\, \tilde{p}_n(1/z) + B_{n+1} \, z^{-n} \, p_n(z).
\ee
These are called Szeg\H{o} recursion equations in \cite{sim1}.
The inverse Szeg\H{o} recursion equations are given by
\be
(1 -A_{n+1} B_{n+1}) z \, p_n(z) = p_{n+1}(z) - A_{n+1} \, z^{n+1} \, \tilde{p}_{n+1}(1/z),
\ee
\be
(1 - A_{n+1} B_{n+1}) z^{-1} \, \tilde{p}_n(1/z) = \tilde{p}_{n+1}(1/z) 
- B_{n+1} \, z^{-n-1} \, p_{n+1}(z).
\ee
Combining these recursion equations, we have the following three-term relations \cite{sim1}:
\be
A_n\,  p_{n+1}(z) = (A_{n+1} + A_n \, z) p_n(z) - (1-A_n B_n) A_{n+1} \, z \, p_{n-1}(z),
\ee
\be
B_n \, \tilde{p}_{n+1}(1/z) = ( B_{n+1} + B_n \, z^{-1}) \tilde{p}_n(1/z) 
- (1 - A_n B_n) B_{n+1} \, z^{-1} \, \tilde{p}_{n-1}(1/z).
\ee

\subsection{String equations}

Recall that
\be
\de \mu(z) = \frac{\de z}{2\pi \im \, z} \exp\Bigl( W_U(z) \Bigr).
\ee
Using the following constraints for $k \in \mathbb{Z}$ and $\ell, m \geq 0$,
\bel{streq0s}
\begin{split}
0 &= \int \de z \frac{\partial}{\partial z}
\left[ \frac{z^k}{2\pi \im} \exp\Bigl( W_U(z) \Bigr) p_{\ell}(z) \tilde{p}_m(1/z) \right] \cr
&=  \int \de \mu(z) \, z^{k+1}\, W_U'(z) p_{\ell}(z) \tilde{p}_m(1/z) 
+ \int \de \mu(z)\, z \frac{\partial}{\partial z} \bigl( p_{\ell}(z) z^{k} \tilde{p}_m(1/z) \bigr),
\end{split}
\ee
we can obtain various polynomial equations for $A_n$ and $B_n$.

In particular, let us consider the following three cases of \eqref{streq0s}: 
(i) $(k, \ell, m)=(-1, n,n-1)$, 
(ii) $(k, \ell, m)=(0,n,n)$
and (iii) $(k, \ell, m) = (1, n-1, n)$. 
They lead to the ``string equations''
\begin{align}
\label{unistr}
\int \de \mu(z)  W_U'(z) p_{n}(z) \tilde{p}_{n-1}(1/z)
&= n( h_n - h_{n-1}),
\\ \label{unistr2}
\int \de \mu(z) \, z\,  W_U'(z) p_{n}(z) \tilde{p}_{n}(1/z)
&=0,
\\ \label{unistr3}
\int \de \mu(z)  \, z^2\, W_U'(z) p_{n-1}(z) \tilde{p}_{n}(1/z)
&= - n( h_n - h_{n-1}).
\end{align}
Here we have used
\begin{align}
\label{strcnstm1s}
\int \de \mu(z) z \frac{\partial}{\partial z} \Bigl( p_{n}(z) z^{-1} \tilde{p}_{n-1}(1/z) \Bigr)
&= - n ( h_n - h_{n-1}),
\\
\label{strcnst0s}
\int \de \mu(z) z \frac{\partial}{\partial z} \Bigl( p_{n}(z)  \tilde{p}_{n}(1/z) \Bigr) &=0,
\\
\label{strcnst1s}
\int \de \mu(z) z \frac{\partial}{\partial z} \Bigl( p_{n-1}(z) z \tilde{p}_{n}(1/z) \Bigr) 
&= n( h_n - h_{n-1}).
\end{align}
For derivation of these equations, see Appendix \ref{DSE1}.

\section{Unitary matrix model with logarithmic potential}

\label{UMMwLog}

Let us consider the unitary matrix model with the following potential\footnote{The matrix model with this potential was considered in \cite{EM10} where the integrations over
the eigenvalues are taken along the real axis.}
\bel{unitpot}
W_U(z) = - \frac{1}{2\, \underline{g}{}_s} \left( z + \frac{1}{z} \right) + M \log z.
\ee
This model is studied in \cite{min9110,his9611,AMM1610}. 
The connection with the Painlev\'{e} III equation is shown
in \cite{his9611}.
In the gauge theory parameters, $\underline{g}{}_s =g_s/\Lambda_2$
and $M=\alpha_{1+2} + N = ( m_2-m_1)/g_s$. See Appendix \ref{CBandIL} for our notation.
We assume that $M$ is an integer. 
Note that $1/(2 \underline{g}{}_s)=\Lambda_2/(2 g_s) = q_{02}$.

When $M=0$, this is the Gross-Witten-Wadia model \cite{GW80,wad1212,wad80}\footnote{
See, for example, \cite{oku1705} and references therein.
It is also known that the GWW model has connection with the Painlev\'{e} III equation \cite{his9609,FW}.}.
In this paper, we think $M$ as a finite parameter. Then, in the large $N$ limit,
the planar free energy of this model does not depend on $M$
and equals  that of the GWW model. Hence this model in the large $N$
also has a third order phase transition at $\widetilde{S}:=N\, \underline{g}{}_s = 1$\footnote{
Another type of generalization of the GWW model can be found, for example, in \cite{BH}.
The phase transition of the asymmetric GWW model is studied in \cite{his9705}.
The GWW model with boundary terms is examined in \cite{min91a, min91b}.}.

\subsection{Moments and related quantities}

The moments of this potential are given by
\be
\begin{split}
\mu_n &= \oint \frac{\de z}{2\pi \im z} \exp\left( - \frac{1}{2\, \underline{g}{}_s}
\left( z + \frac{1}{z} \right) \right) z^{M+n} \cr
&= \left( -\frac{1}{2 \, \underline{g}{}_s} \right)^{|M+n|}
\sum_{k=0}^{\infty} \frac{1}{k!\,  (k+|M+n|)!} \left( \frac{1}{2\, \underline{g}{}_s} \right)^{2k} 
=(-1)^{M+n} I_{|M+n|}(1/\underline{g}{}_s),
\end{split}
\ee
where $I_{\nu}(z)$ is the modified Bessel function of the first kind:
\be
I_{\nu}(z) = \left( \frac{z}{2} \right)^{\nu} \sum_{k=0}^{\infty}
\frac{1}{k!\, \Gamma(\nu+k+1)} \left( \frac{z}{2} \right)^{2k}.
\ee
Note that
\be
\begin{split}
\mathcal{K}^{(n)}_k &=\det ( \mu_{j-i+k} )_{1 \leq i,j \leq n} \cr
&= \det \Bigl( (-1)^{M+j-i+k} I_{|M+j-i+k|}(1/\underline{g}{}_s) \Bigr)_{1 \leq i,j \leq n} \cr
&=(-1)^{n(M+k)} K^{(n)}_{M+k},
\end{split}
\ee
where
\bel{Knn}
K^{(n)}_{\nu} := \det\Bigl( I_{j-i+\nu}(1/\underline{g}{}_s)  \Bigr)_{1 \leq i,j \leq n}, \qq
( \nu \in \mathbb{C};  n=0,1,2,\dotsm).
\ee
For an integer $k$, 
it holds that $I_{-k}(z) = I_k(z)$. Therefore, for $j-i+M+k \in \mathbb{Z}$, we have
$I_{j-i+M+k}(1/\underline{g}{}_s)=I_{|j-i+M+k|}(1/\underline{g}{}_s)$.
Also, we have
$K^{(n)}_{-k} = K^{(n)}_k$ ($k \in \mathbb{Z}$). 
For later convenience, we have defined $K^{(n)}_{\nu}$ \eqref{Knn} as a determinant of $I_{j-i+\nu}(1/\underline{g}{}_s)$
such that the index $M$ in $K^{(n)}_{M+k}$ can be analytically continued from an integer to any complex number.

Note that
\be
\tau_n = \mathcal{K}^{(n)}_0 = (-1)^{nM} K^{(n)}_M.
\ee
The normalization constants of the orthogonal polynomials
are given by 
\be
h_n = \frac{\mathcal{K}_0^{(n+1)}}{\mathcal{K}_0^{(n)}}
=(-1)^M \frac{K^{(n+1)}_M}{K^{(n)}_M}.
\ee
In particular,  $h_0=(-1)^M \, I_{M}(1/\underline{g}{}_s)$.  

The constant terms of the orthogonal polynomials are written in terms of $K^{(n)}_k$ as follows:
\be
A_n =p_n(0) = (-1)^n \frac{\mathcal{K}^{(n)}_1}{\mathcal{K}^{(n)}_0}=
\frac{K^{(n)}_{M+1}}{K^{(n)}_M},
\ee
\be
B_n = \tilde{p}_n(0) = (-1)^n \frac{\mathcal{K}^{(n)}_{-1}}{\mathcal{K}^{(n)}_0}
=\frac{K^{(n)}_{M-1}}{K^{(n)}_M}.
\ee

The partition function \eqref{uZUN} can be written in terms of these objects:
\bel{PartZN2}
\underline{Z}_{U(N)} = (-1)^{MN} K^{(N)}_M=
\prod_{k=0}^{N-1} h_k = h_0^N \prod_{j=1}^{N-1} \bigl(1 - A_j B_j \bigr)^{N-j}.
\ee

\subsection{The partition function as the tau function of PIII${}'$}

The partition function \eqref{PartZN2} 
is essentially the tau function of the Painlev\'{e} III${}'$ equation\footnote{For the appearance of PIII in Ising system, see \cite{WMTB, MPW, Mc90}. }.

For $q=q(s)$ and $p=p(s)$, let us choose
the Hamiltonian for the PIII${}'$ as
\be
H_{\mathrm{III}'}(s)
= \frac{1}{s} \left[
q^2 \, p^2
- (q^2 + v_2\, q - s) p + \frac{1}{2} (v_1+v_2) q \right].
\ee
The Hamiltonian system
\be
\frac{\de q}{\de s} = \frac{\partial H_{\mathrm{III}'}}{\partial p},
\qq
\frac{\de p}{\de s} = - \frac{\partial H_{\mathrm{III}'}}{\partial q}
\ee
is equivalent to the PIII${}'$ equation
\be
\frac{\de^2 q}{\de s^2}
= \frac{1}{q} \left( \frac{\de q}{\de s} \right)^2
- \frac{1}{s} \left( \frac{\de q}{\de s} \right)
+ \frac{q^2}{4\, s^2} ( \gamma\, q + \alpha)
+ \frac{\beta}{4\, s} + \frac{\delta}{4\, q},
\ee
with
\be
\alpha = - 4\, v_1, \qq
\beta = 4(v_2+1), \qq
\gamma=4, \qq
\delta=-4.
\ee
The tau function $\tau(s) = \tau_{\mathrm{III}'}(s)$ for PIII${}'$ is introduced by
\be
H_{\mathrm{III}'}(s) = \frac{\de}{\de s} \log \tau(s).
\ee
Let
\be
\sigma(s):=s H_{\mathrm{III}'}(s) = q^2 \, p^2
- (q^2 + v_2\, q - s) p + \frac{1}{2} (v_1+v_2) 
= s \frac{\de}{\de s} \log \tau(s).
\ee
This function satisfies the $\sigma$-form of PIII${}'$:
\be
( s \sigma'')^2 - 4 ( \sigma - s \sigma') \sigma'(\sigma'-1)
- \left( v_2 \, \sigma' - \frac{1}{2} ( v_1 + v_2) \right)^2 = 0.
\ee
Combining eq.(4.8) of \cite{FW02} and Corollary 4.5 of \cite{FW02}\footnote{
The referee of our previous paper \cite{IOYanok1}  called our attention to
this Corollary 4.5 of \cite{FW02}.}
, we can show that
\be
\tau(s) = s^{(1/2)MN} K^{(N)}_M(2\sqrt{s})
\ee
is the tau function of PIII${}'$ with
\be
v_1 = M+N= - \frac{2m_1}{g_s}, \qq v_2 = - M+N= - \frac{2m_2}{g_s}.
\ee
Hence, if we set
\be
2 \sqrt{s} = \frac{1}{\underline{g}{}_s}, \qq
s = \frac{1}{4\, \underline{g}{}_s{}^2},
\ee
the partition function $\underline{Z}{}_{U(N)}(M) = (-1)^{MN} K^{(N)}_M$
is the tau function of PIII${}'$ (up to an overall factor $s^{(1/2)MN}$).

It is well-known that PIII${}'$ equation is equivalent to the PIII equation
\be
\frac{\de^2 y}{\de \mathfrak{t}^2}
= \frac{1}{y} \left( \frac{\de y}{\de \mathfrak{t}} \right)^2
- \frac{1}{\mathfrak{t}} \frac{\de y}{\de \mathfrak{t}}
+ \frac{1}{\mathfrak{t}} 
\Bigl( \alpha  y^2 + \beta \Bigr) + \gamma y^3 + \frac{\delta}{y}.
\ee
The corresponding Hamiltonian for $y=y(\mathfrak{t})$ and its conjugate momentum 
$p_y=p_y(\mathfrak{t})$ is given by
\be
H_{\mathrm{III}}
= \frac{1}{\mathfrak{t}} 
\Bigl[ 2\, y^2\, p_y^2 
- \bigl\{ 2\, \mathfrak{t}\, y^2 + (2 v_2+1)y - 2\, \mathfrak{t} \bigr\} p_y
+ (v_1+v_2) \mathfrak{t}\, y \Bigr].
\ee
The variables and the Hamiltonian of PIII${}'$ are related to those of PIII by 
\be
q = \mathfrak{t}\, y, \qq p = \frac{p_y}{\mathfrak{t}}, \qq s =\mathfrak{t}^2,
\qq
H_{\mathrm{III}'}
= \frac{1}{2\, \mathfrak{t}} \left( H_{\mathrm{III}} + \frac{y \, p_y}{\mathfrak{t}} \right).
\ee
Hence, we find the connection between  the coupling constant $\underline{g}{}_s$ and
the time variable $\mathfrak{t}$ of PIII equation: 
$\mathfrak{t} = 1/(2 \underline{g}{}_s)$.

The B\"{a}cklund transformations of the PIII${}_1$ (or PIII($D_6^{(1)}$) ) form the affine Weyl group of type $(2A_1)^{(1)}$.
The translation subgroup generates the alt-dPII equation. The translations lead to integer shifts of parameters $v_{1,2}$. 
In terms of the gauge theory parameters, 
they correspond to constant shifts of mass parameters $m_{1,2}$.

\subsection{String equations}

Let us write the string equations \eqref{unistr}, \eqref{unistr2} and \eqref{unistr3}
explicitly for the case of the potential \eqref{unitpot}. Since
\be
W'_U(z) = - \frac{1}{2\, \underline{g}{}_s} \left( 1 - \frac{1}{z^2} \right) + \frac{M}{z},
\ee
we have
\bel{USE1}
\int \de \mu(z)\, W'_U(z) p_n(z) \tilde{p}_{n-1}(1/z)
=\frac{1}{2\, \underline{g}{}_s} \bigl( \widetilde{C}^{(n)}_n + \widetilde{C}^{(n-1)}_{n-1} \bigr) h_n + M h_n,
\ee
\bel{USE2}
\int \de \mu(z)\, z\, W'_U(z) p_n(z) \tilde{p}_n(1/z)
=- \frac{1}{2\, \underline{g}{}_s} \bigl( C^{(n)}_n - \widetilde{C}^{(n)}_n \bigr) h_n + M h_n,
\ee
\bel{USE3}
\int \de \mu(z)\, z^2\, W'_U(z) p_{n-1}(z) \tilde{p}_n(1/z)
=- \frac{1}{2\, \underline{g}{}_s} \bigl( C^{(n)}_n + C^{(n-1)}_{n-1} \bigr) h_n + M h_n.
\ee
Here we have used \eqref{recp1a} and \eqref{recp2a}. (See Appendix \ref{DSE2} for details.)

Then the string equations \eqref{unistr}, \eqref{unistr2}, \eqref{unistr3} for this potential become
\be
\begin{split}
\frac{1}{2\, \underline{g}{}_s}\bigl( \widetilde{C}^{(n)}_n + \widetilde{C}^{(n-1)}_{n-1} \bigr)  + M 
&= n \left( 1 - \frac{h_{n-1}}{h_n} \right), \cr
- \frac{1}{2\, \underline{g}{}_s} \bigl( C^{(n)}_n - \widetilde{C}^{(n)}_n \bigr)  + M 
&= 0, \cr
- \frac{1}{2\, \underline{g}{}_s} \bigl( C^{(n)}_n + C^{(n-1)}_{n-1} \bigr)  + M 
&= - n \left( 1- \frac{h_{n-1}}{h_n} \right).
\end{split}
\ee
Using
\be
\frac{h_n}{h_{n-1}} = 1 - A_n B_n,
\qq
C^{(n)}_n = - A_{n+1} B_n, \qq
\widetilde{C}^{(n)}_n = - A_n B_{n+1},
\ee
the string equations lead to
the following recursion relations for  $A_n$ and $B_n$:
\bel{ABrec}
A_{n+1} = - A_{n-1} + \frac{2\, n \underline{g}{}_s\, A_n}{1 - A_n\, B_n}, \qq
B_{n+1} = - B_{n-1} + \frac{2\, n \underline{g}{}_s\, B_n}{1- A_n\, B_n}, 
\ee
\bel{ABrec2}
A_n B_{n+1} - A_{n+1} B_n = 2\, M\, \underline{g}{}_s.
\ee
With the initial conditions $A_0=B_0=1$, and
\bel{initAB}
A_1 = \frac{I_{M+1}(1/\underline{g}{}_s)}{I_{M}(1/\underline{g}{}_s)}, \qq
B_1 = \frac{I_{M-1}(1/\underline{g}{}_s)}{I_{M}(1/\underline{g}{}_s)},
\ee
the remaining constants $A_n$ and $B_n$ are characterized by the recursion relations \eqref{ABrec}, \eqref{ABrec2}.
We remark that one of recursion relations \eqref{ABrec} can be obtained 
by combining the other of \eqref{ABrec} with \eqref{ABrec2}.

Recall that the modified Bessel function satisfies the following recursion relation:
\be
I_{\nu-1}(z) - I_{\nu+1}(z) = (2\nu/z) I_{\nu}(z).
\ee
By examining \eqref{ABrec2} for $n=0$, we can see that
the range of the parameter $M$ in the initial conditions \eqref{initAB} can be extended from 
the integers to any complex numbers. Furthermore,
\be
A_n(M) = \frac{K^{(n)}_{M+1}}{K^{(n)}_M}, \qq
B_n(M) = \frac{K^{(n)}_{M-1}}{K^{(n)}_M}, \qq (M \in \mathbb{C})
\ee
indeed solve the string equations \eqref{ABrec} and \eqref{ABrec2}. Here $K^{(n)}_{\nu}$
is defined by \eqref{Knn}.

\subsubsection{Singularity confinement}

We show that the string equations \eqref{ABrec} and \eqref{ABrec2} pass the ``singularity confinement criterion''
\cite{GRP91} which may be the discrete version of the Painlev\'{e} property.

The string equation \eqref{ABrec} develops a singularity when $A_n B_n=1$.
Let us assume that this occurs for $n=n_0$.
Using the invariance of the recursion relations under
\be
A_k \rightarrow \lambda A_k, \qq
B_k \rightarrow \lambda^{-1} B_k,
\ee
we consider the case
\be
A_{n_0} = 1 + \varepsilon A_{n_0-1}, \qq
B_{n_0} = 1 + \varepsilon B_{n_0-1},
\ee
where $\varepsilon$ is a small parameter.
The ``initial data'' $A_{n_0-1}$ and $B_{n_0-1}$ are finite and obey the relation
\be
A_{n_0-1} - B_{n_0-1} = 2\, M\, \underline{g}{}_s.
\ee
For simplicity we assume that $A_{n_0-1}+B_{n_0-1} \neq 0$. Then the singularities appear at $n=n_0+1$:
\be
A_{n_0+1} =  -\frac{2\, n_0 \, \underline{g}{}_s}{(A_{n_0-1} + B_{n_0-1}) \varepsilon}
-A_{n_0-1} - \frac{2 \, n_0 \, A_{n_0-1}^2 \, \underline{g}{}_s}{(A_{n_0-1}+B_{n_0-1})^2} + O(\varepsilon),
\ee
\be
B_{n_0+1} = - \frac{2\, n_0 \, \underline{g}{}_s}{(A_{n_0-1}+B_{n_0-1}) \varepsilon}
-B_{n_0-1} - \frac{2 \, n_0 \, B_{n_0-1}^2 \, \underline{g}{}_s}{(A_{n_0-1}+B_{n_0-1})^2} + O(\varepsilon).
\ee
At $n=n_0+2$, 
\be
A_{n_0+2} = - 1 + \frac{A_{n_0-1}+(n_0+1) B_{n_0-1}}{n_0} \varepsilon + O(\varepsilon^2),
\ee
\be
B_{n_0+2} = -1 + \frac{(n_0+1)A_{n_0-1}+B_{n_0-1}}{n_0} \varepsilon + O(\varepsilon^2),
\ee
hence it again holds that $A_{n_0+2} B_{n_0+2}=1 + O(\varepsilon)$, 
but there is no singularity at the next level $n=n_0+3$:
\be
A_{n_0+3} = \frac{1}{n_0+2} \Bigl( 2 (n_0+1) \underline{g}{}_s + A_{n_0-1} - (n_0+1) B_{n_0-1} \Bigr)+O(\varepsilon),
\ee
\be
B_{n_0+3} = \frac{1}{n_0+2} \Bigl( 2 (n_0+1) \underline{g}{}_s -(n_0+1) A_{n_0-1} + B_{n_0-1} \Bigr)+O(\varepsilon).
\ee
The singularity is confined to the level $n=n_0+1$ and the memory of the initial data is kept in crossing this singularity.
Therefore, the string equations \eqref{ABrec} and \eqref{ABrec2} pass the criterion. 

We remark that if we replace the recursion relations \eqref{ABrec} by
\be
A_{n+1} = - A_{n-1} + \frac{ (a n + b) A_n}{1-A_n B_n}, \qq
B_{n+1} = - B_{n-1} + \frac{( a n + b) B_n}{1-A_n B_n}
\ee
where $a$ and $b$ are arbitrary constants, these equations with \eqref{ABrec2} also pass the singularity confinement
test.

\subsubsection{Equations for $R_n^2=A_n B_n$}

Note that the partition function \eqref{PartZN2} depends on $A_j$ and $B_j$ only through 
their product $A_j B_j$. 
Let $A_n = R_n D_n$ and $B_n = R_n/D_n$. Then the partition function  \eqref{PartZN2} becomes
\bel{PartZR}
\underline{Z}_{U(N)} = h_0^N \prod_{j=1}^{N-1} ( 1 - R_j^2 )^{N-j}.
\ee

The  equation \eqref{ABrec2} turns into
\be
R_n R_{n+1} \left( \frac{D_n}{D_{n+1}} - \frac{D_{n+1}}{D_n} \right) = 2 \, M \, \underline{g}{}_s.
\ee
This leads to
\be
\frac{D_n}{D_{n+1}} = \frac{\dis M \, \underline{g}{}_s 
+ \sqrt{ R_n^2 \, R_{n+1}^2 + M^2 \, \underline{g}{}_s^2}}{R_n \, R_{n+1}},
\ee
\be
\frac{D_{n+1}}{D_{n}} = \frac{\dis - M \, \underline{g}{}_s
+ \sqrt{ R_n^2 \, R_{n+1}^2 +  M^2 \, \underline{g}{}_s^2}}{R_n \, R_{n+1}}.
\ee

By substituting these relations into the remaining relations \eqref{ABrec}, we find
\bel{Rrec}
(1 - R_n^2) \Bigl( \sqrt{R_n^2 \, R_{n+1}^2 + M^2 \, \underline{g}{}_s^2 }
+ \sqrt{R_n^2\, R_{n-1}^2 + M^2 \, \underline{g}{}_s^2 } \Bigr) = 2\, n \, \underline{g}{}_s\, R_n^2.
\ee
This is equivalent to
\bel{STREQ}
\begin{split}
0=& \eta_n^2 \Bigl[ 
\xi_n^2 (1 - \xi_n)^2  - \eta_n^2 \, \xi_n^2 + \zeta^2 (1 - \xi_n)^2
\Bigr] \cr 
& +\frac{1}{2} \, \eta_n^2 \, \xi_n\, ( 1 - \xi_n)^2 (\xi_{n+1} - 2\, \xi_n + \xi_{n-1}) 
 -  \frac{1}{16} (1 - \xi_n)^4 ( \xi_{n+1} - \xi_{n-1})^2,
\end{split}
\ee
where $\xi_n \equiv R_n^2$, $\eta_n \equiv n \, \underline{g}{}_s$, $\zeta \equiv M\, \underline{g}{}_s$.

When $M=0$ (i.e., with no logarithmic potential), \eqref{Rrec} reduces to the string equation 
considered in \cite{PS90a} 
\be
(1 - R_n^2) \, R_n ( R_{n+1} + R_{n-1} ) = 2 \, n \, \underline{g}{}_s \, R_n^2.
\ee

\subsubsection{Relation with alternate discrete Painlev\'{e} II equation}

Let us introduce variables $x_n$ and $y_n$ by
\be
x_n:= \frac{A_{n+1}}{A_n}, \qq y_n:= \frac{B_{n+1}}{B_n}, \qq
(n=0,1,2,\dotsm).
\ee
They respectively obey the alternate discrete Painlev\'{e} II equation \cite{FGR93,NSKGR96}
with different values of the parameter $\tilde{\mu}$. 
(See Appendix \ref{altdPII} for details.)
With the initial conditions $A_0=1$ and $B_0=1$, $A_n$ and $B_n$ can be expressed by these variables:
\be
A_n=\prod_{k=0}^{n-1} x_k, \qq
B_n=\prod_{k=0}^{n-1} y_k.
\ee

\subsection{Small $\underline{g}{}_s$ expansion}

In this subsection, we comment on a perturbative solution for the string equations \eqref{ABrec}, \eqref{ABrec2}. 
We consider the perturbation in $\underline{g}{}_s$. We remark that in the time variable $\mathfrak{t}$ 
\bel{timeIII}
\mathfrak{t} = \frac{1}{2\, \underline{g}{}_s} =\frac{\Lambda_2}{2\, g_s} = 2\, q_{02}
\ee
for the Painlev\'{e} III${}_1$, the small $\underline{g}{}_s$ corresponds to the large $\mathfrak{t}$ regime.

Recall that $A_n(M)$ and $B_n(M)$ are expressed by the modified Bessel functions $I_{\nu}(1/\underline{g}{}_s)$.
The large $z$ behavior of the modified Bessel function is given by
\be
I_{\nu}(z) \sim \frac{\ex^z}{\sqrt{2\pi z}}
\sum_{k=0}^{\infty} (-1)^k \frac{\widetilde{\mathcal{C}}_k(\nu)}{k!\, (2z)^k} +
\frac{\ex^{-z+(\nu+(1/2)) \pi \im}}{\sqrt{2\pi z}}
\sum_{k=0}^{\infty} \frac{\widetilde{\mathcal{C}}_k(\nu)}{k!\, (2z)^k},
\ee
for $-\pi/2 < \mathrm{arg}\, z < 3\pi/2$. Here
\bel{tCkM}
\widetilde{\mathcal{C}}_k( \nu) := \frac{\Gamma(\nu+k+(1/2))}{\, \Gamma(\nu - k + (1/2))}
= \prod_{j=1}^k \bigl( \nu^2 - (j-1/2)^2 \bigr).
\ee
Note that $\widetilde{\mathcal{C}}_k(-\nu) = \widetilde{\mathcal{C}}_k(\nu)$.

We assume that $\mathrm{Re}\, \underline{g}{}_s>0$.
Then the small $\underline{g}{}_s$ asymptotic expansion yields
\bel{asymptnu}
I_{\nu}(1/\underline{g}{}_s)  
\sim \left( \frac{\underline{g}{}_s}{2\pi} \right)^{1/2} \ex^{1/\underline{g}{}_s} 
\left[
\sum_{k=0}^{\infty}\frac{(-1)^k}{k!}\bigl(1 + 
\ex^{(\nu+k+(1/2))\pi \im}\, 
\ex^{-2/\underline{g}{}_s} \bigr)
 \widetilde{\mathcal{C}}_k(\nu) \left( \frac{\underline{g}{}_s}{2} \right)^k
 \right].
\ee

\subsubsection{Perturbative contribution}

The asymptotic expansion \eqref{asymptnu} has 
a part which is exponentially small term.
We interpret that this part yields the ``instanton contribution''.
As an approximation, we omit this instanton part:
\be
I_{\nu}(1/\underline{g}{}_s) \sim \left( \frac{\underline{g}{}_s}{2\pi} \right)^{1/2} 
\ex^{1/\underline{g}{}_s} 
\sum_{k=0}^{\infty}\frac{(-1)^k}{k!}
 \widetilde{\mathcal{C}}_k(\nu) \left( \frac{\underline{g}{}_s}{2} \right)^k.
\ee
With the truncated initial conditions
\be
A_0(M) = B_0(M) = 1,
\ee
\bel{A1}
A_1(M) = \frac{\dis \sum_{k=0}^{\infty} \frac{(-1)^k }{k!} \widetilde{\mathcal{C}}_k(M+1) 
\left( \frac{\underline{g}{}_s}{2} \right)^k}
{\dis \sum_{k=0}^{\infty}\frac{(-1)^k}{k!} \widetilde{\mathcal{C}}_k(M) 
\left( \frac{\underline{g}{}_s}{2}
\right)^k},
\ee
\bel{B1}
B_1(M) = \frac{\dis \sum_{k=0}^{\infty} \frac{(-1)^k }{k!} \widetilde{\mathcal{C}}_k(M-1) 
\left( \frac{\underline{g}{}_s}{2}
\right)^k}
{\dis \sum_{k=0}^{\infty}\frac{(-1)^k}{k!} \widetilde{\mathcal{C}}_k(M) 
\left( \frac{\underline{g}{}_s}{2}
\right)^k},
\ee
we can solve the string equations and determine $A_n$ and $B_n$ as the power series in 
$\underline{g}{}_s$. Using
\be
\widetilde{\mathcal{C}}_{k+1}(M+1) - \widetilde{\mathcal{C}}_{k+1}(M-1) = 4 \, M (k+1) \, 
\widetilde{\mathcal{C}}_{k}(M),
\ee
we can check that the above truncated initial conditions are consistent with
\be
A_0(M) B_1(M) - A_1(M) B_0(M) = B_1(M) - A_1(M) = 2 \, M \underline{g}{}_s.
\ee
Note that for the perturbative case, it holds that $B_1(M)=A_1(-M)$ and 
subsequently $B_n(M)=A_n(-M)$. 
(When the ``instanton contributions'' are taken into account, $B_n(M)=A_n(-M)$ does not hold for generic $M$.)

We write the coefficients in the $\underline{g}{}_s$ expansion as follows:
\bel{Anexp}
A_n(M) = \sum_{k=0}^{\infty} \frac{A_{n,k}(M)}{k!} \left( \frac{\underline{g}{}_s}{2} \right)^k,
\qq
B_n(M) = \sum_{k=0}^{\infty} \frac{A_{n,k}(-M)}{k!} \left( \frac{\underline{g}{}_s}{2} \right)^k.
\ee
We have determined the coefficients $A_{n,k}(M)$ for $0 \leq k \leq 11$. First few of them are
\bel{An4}
\begin{split}
A_{n,0}(M) &=1, \cr
A_{n,1}(M) &= - n(2M+1), \cr
A_{n,2}(M) &= n^2 (2M+1)(2M-1), \cr
A_{n,3}(M) &= - n(2M+1)(2M-1)\bigl\{ n^2(2M-3) - (2M+3) \bigr\}, \cr
A_{n,4}(M) &= n^2(2M+1)(2M-1)(2M-5)\bigl\{ n^2(2M-3) -4(2M+3) \bigr\}.
\end{split}
\ee
Explicit form of $A_{n,k}(M)$ for $5 \leq k \leq 11$ are given in Appendix \ref{AnkM}.

Then, we find the small $\underline{g}{}_s$ expansion of $R_n^2(M)$ and $D_n^2(M)$:
\bel{ABpert}
\begin{split}
R_n^2(M) &= A_n(M) B_n(M) = A_n(M) A_n(-M) \cr
&=
1 -  n \underline{g}{}_s 
+ \frac{1}{8} n \bigl( (2M)^2-1 \bigr) \underline{g}{}_s^3 
+ \frac{1}{4} n^2 \bigl((2M)^2-1\bigr) \underline{g}{}_s^4 \cr
& +\frac{3}{128} n \bigl( (2M)^2-1 \bigr)\bigl\{ 16\, n^2 - \bigl((2M)^2-3^2\bigr) \bigr\} \underline{g}{}_s^5 \cr
& + \frac{1}{8} \, n^2 \bigl((2M)^2-1 \bigr) \bigl\{ 4\, n^2 - \bigl((2M)^2-3^2 \bigr) \bigr\} 
\underline{g}{}_s^6 \cr
& + \frac{5}{1024} n \bigl( (2M)^2-1) \Bigl\{
128\, n^4 - 80 \bigl( (2M)^2-3^2 \bigr) n^2
+ \bigl( (2M)^2-3^2 \bigr) \bigl( (2M)^2-5^2 \bigr)
\Bigr\}
\underline{g}{}_s^7 \cr
&+\frac{3}{16}n^2 \bigl( (2M)^2-1 \bigr)
\Bigl\{
4\, n^4 - 5 \bigl( (2M)^2-3^2 \bigr) n^2 + \bigl( (2M)^2-3^2 \bigr)(M^2-6) 
\Bigr\}
\underline{g}{}_s^8 \cr
&+ 7 \, n \bigl( (2M)^2-1 \bigr) \left\{
\frac{1}{8} n^6 - \frac{35}{128} \bigl( (2M)^2-3^2 \bigr) n^4 \right. \cr
& \qq \qq \qq \qq \left.
+ \frac{7}{1024} \bigl( (2M)^2 - 3^2 \bigr) \bigl( 20\, M^2-117 \bigr)n^2 \right. \cr
&  \qq \qq \qq \qq \left. - \frac{5}{32768} \bigl( (2M)^2-3^2 \bigr)\bigl( (2M)^2-5^2 \bigr) \bigl( (2M)^2-7^2 \bigr)
\right\}\underline{g}{}_s^9 \cr
&+O(\underline{g}{}_s^{10}).
\end{split}
\ee
\be
\begin{split}
D_n^2(M) &= \frac{A_n(M)}{B_n(M)} \cr
&= 1 - 2 \, n \, M \, \underline{g}{}_s
+  n^2\, M (2 M-1) \underline{g}{}_s^2 \cr
&-\frac{1}{12} M(2M-1) n \bigl\{ 8 (M-1)n^2 - (2M+1) \bigr\}
\underline{g}{}_s^3 \cr
&+ \frac{1}{12} M (2M-1)(2M-3)n^2 
\bigl\{ 2(M-1) n^2 - (2M+1) \bigr\}
\underline{g}{}_s^4 \cr
&+O(\underline{g}{}_s^5).
\end{split}
\ee
We know the exact expressions of $R_n^2(M)$ and $D_n^2(M)$ up to order $\underline{g}{}_s^{11}$ terms, but they
are too lengthy. So we truncated them to the above orders.

\subsubsection{Properties of $R_n^2(M)$}

By inspecting the small $\underline{g}{}_s$ expansion \eqref{ABpert},
we may parametrize the $M$ dependence of $R_n^2(M)$ as follows:
\bel{rnstr}
\begin{split}
R_n^2(M) &= \sum_{k=0}^{\infty}  \widetilde{\mathcal{C}}_k(M) r_{n,k} \cr
&= r_{n,0} + (M^2-(1/2)^2) r_{n,1} + (M^2 -(1/2)^2)(M^2-(3/2)^2) r_{n,2} \cr
&+ (M^2 - (1/2)^2) ( M^2 - (3/2)^2) (M^2 - (5/2)^2) r_{n,3} + \dotsm,
\end{split}
\ee
where $r_{n,k}$ are independent on $M$ and $\widetilde{\mathcal{C}}_k(M)$ are introduced in \eqref{tCkM}.

In the string equation, we can think $M$ as a free parameter. 
Hence it can be set to a half-integer.
From \eqref{rnstr}, we have following conditions:
\be
\begin{split}
R_n^2(1/2) &= A_n(1/2) A_n(-1/2) = r_{n,0}, \cr
R_n^2(3/2) &= A_n(3/2) A_n(-3/2) = r_{n,0} + 2\, r_{n,1}, \cr
R_n^2(5/2) &= A_n(5/2) A_n(-5/2) = r_{n,0} + 6\, r_{n,1} + 24 \, r_{n,2},
\end{split}
\ee
etc.  In general, we have
\be
R_n^2(j+1/2) = \sum_{k=0}^j \widetilde{\mathcal{C}}_k(j+1/2) r_{n,k}, \qq
(j=0,1,2,\dotsm).
\ee
By solving these equations for $r_{n,k}$, we can express $r_{n,k}$ as linear combination of $R_n^2(j+1/2)$'s. For example,
\be
\begin{split}
r_{n,0} &= R_n^2(1/2), \cr
r_{n,1} &= \frac{1}{2}(- R_n^2(1/2)+ R_n^2(3/2) ), \cr
r_{n,2} &= \frac{1}{24}(2\, R_n^2(1/2) -3\, R_n^2(3/2) + R_n^2(5/2) ).
\end{split}
\ee
The general solutions are given by
\be
r_{n,k} = \sum_{j=0}^k (-1)^{k-j} \frac{(2j+1)}{(k-j)!\, (k+j+1)!} R_n^2(j+1/2).
\ee
By substituting this result into \eqref{rnstr}, we can express $R_n^2(M)$ in terms of 
$R_n^2(j+1/2)$'s:
\be
R_n^2(M) = \sum_{j=0}^{\infty} \widetilde{J}_j(M) \, R_n^2(j+1/2),
\ee
where
\bel{JjMdef}
\widetilde{J}_j(M) = (2j+1) \sum_{k=0}^{\infty}
 \frac{(-1)^k \, \widetilde{\mathcal{C}}_{j+k}(M)}{ k! \, (k+2j+1)!}.
\ee
In Appendix \ref{JjM}, we show that $\widetilde{J}_j(M)$ is simplified to
\bel{JjMcos}
\widetilde{J}_j(M) = \frac{(-1)^{j+1} (2j+1) \cos(\pi M)}{\pi\,  ( M^2 - (j+1/2)^2) }.
\ee
Note that
\be
\lim_{M \rightarrow j+1/2} \widetilde{J}_j(M) =1.
\ee
Now we have the following expansion:
\bel{RnM2}
R_n^2(M) = \frac{1}{\pi} \cos (\pi M) \left(
\sum_{j=0}^{\infty} \frac{(-1)^j (2j+1)}{\{ (j+1/2)^2 - M^2 \}} \, R_n^2(j+1/2) \right).
\ee

When $M$ is a half-integer, the initial function $A_1$ \eqref{A1} and $B_1$ \eqref{B1} 
become rational functions of
$\underline{g}{}_s$, hence
$A_n=A_n(M)$ and $B_n=A_n(-M)$ also become rational functions. 
For example,
\begin{align*}
A_n(-1/2) &=1,& A_n(1/2) &= 1 - n\, \underline{g}{}_s, \cr
A_n(-3/2) &= \frac{1}{1 - n\, \underline{g}{}_s},&
A_n(3/2) &= \frac{1 - 3 \, n \, \underline{g}{}_s + 3 \, n^2 \, \underline{g}{}_s^2 - n(n^2-1) \underline{g}{}_s^3}{1 - n\, \underline{g}{}_s},
\end{align*}
etc. Let
\bel{ABV}
A_n(j+1/2) = \frac{1}{V_{n,j+1}}, \qq
B_n(j+1/2) = A_n(-j-1/2) = V_{n,j}, \qq
(j =0,1,2,\dotsm).
\ee
The second equation of \eqref{ABrec} can be rewritten as
\be
A_n = \frac{1}{B_n} - \frac{2\, n \underline{g}{}_s}{B_{n+1} + B_{n-1}}.
\ee
By substituting \eqref{ABV} into this equation, we find
the following recursion relations for $V_{n,j}$:
\be
\frac{1}{V_{n,j+1}} = \frac{1}{V_{n,j}} - \frac{2\, n \underline{g}{}_s}{V_{n+1,j} + V_{n-1,j}}.
\ee
With the initial conditions,
\be
V_{n,0} = 1, \qq
V_{n,1} = \frac{1}{1 - n\, \underline{g}{}_s},
\ee
we can determine $V_{n,j}$. The recursion relation \eqref{ABrec2} yields the following consistency condition:
\be
\frac{V_{n+1,j}}{V_{n,j+1}} - \frac{V_{n,j}}{V_{n+1,j+1}} = (2j+1) \underline{g}{}_s.
\ee 
From this condition, $V_{n,j}$ is seen to have an expression which looks like a continued fraction.
 
For example, for $j=2$, 
\be
\begin{split}
V_{n,2} &= \frac{1}{\dis 1 - n \, \underline{g}{}_s - \frac{2 \, n \underline{g}{}_s}{\dis \frac{1}{1 - (n+1) \underline{g}{}_s}
+ \frac{1}{1 - (n-1) \underline{g}{}_s}}} \cr
&= \frac{1 - n\, \underline{g}{}_s}
{1 - 3 \, n \, \underline{g}{}_s + 3 \, n^2 \, \underline{g}{}_s^2 - n(n^2-1) \underline{g}{}_s^3}.
\end{split}
\ee

We also have the following relations:
\be
R_n^2(j+1/2) = A_n(j+1/2) B_n(j+1/2) = \frac{V_{n,j}}{V_{n,j+1}}.
\ee
For $j=0,1$, the explicit forms of $R_n^2(j+1/2)$ are given by
\be
R_n^2(1/2) = \frac{1}{V_{n,1}} = 1 - n \, \underline{g}{}_s,
\ee
\be
\begin{split}
R_n^2(3/2) &= \frac{V_{n,1}}{V_{n,2}} \cr
&= \frac{(1 - 3 \, n \, \underline{g}{}_s + 3 \, n^2 \, \underline{g}{}_s^2 - n(n^2-1) \underline{g}{}_s^3)}{(1- n \, \underline{g}{}_s)^2}.
\end{split}
\ee
We finally have
\be
R_n^2(M) 
= \frac{1}{\pi} \cos (\pi M) \left(
\sum_{j=0}^{\infty} \frac{(-1)^j (2j+1)}{\{ (j+1/2)^2-M^2 \}} \frac{V_{n,j}}{V_{n,j+1}} \right).
\ee

\subsection{Free energy}

By discarding an irrelevant phase factor $(-1)^{MN}$, let us define the free energy of 
the partition function \eqref{PartZR} by
\bel{UFE}
\begin{split}
\mathcal{F}&:= \log\Bigl( (-1)^{MN} \underline{Z}{}_{U(N)} \Bigr) \cr
&= N \log \Bigl( I_M(1/\underline{g}{}_s) \Bigr)
+ \sum_{j=1}^N (N-j) \log \bigl( 1 - R_j^2(M) \bigr).
\end{split}
\ee
Let us examine the $\underline{g}{}_s$ expansion of this free energy.

From \eqref{ABpert}, we can see that the perturbative contribution of $1 - R_j^2$ is given by
\be
1 - R_j^2(M) = j \underline{g}{}_s
\left( 1 - \frac{1}{2} \bigl(M^2-(1/4) \bigr) \underline{g}{}_s{}^2+ O(\underline{g}{}_s{}^3) \right).
\ee 
Also, by inspecting the asymptotic behavior \eqref{asymptnu}, we can see that it is convenient to rearrange the 
free energy \eqref{UFE} as follows
\bel{Fnpp}
\mathcal{F} = \mathcal{F}^{(\mathrm{G})} + \mathcal{F}^{(\mathrm{n})},
\ee
where
\be
\mathcal{F}^{(\mathrm{G})}
:= \log \left[ \frac{(2\pi \underline{g}{}_s)^{(1/2)N^2}}{\mathrm{vol}(U(N))} \right],
\ee
and
\be
\mathcal{F}^{(\mathrm{n})}
:= \frac{N}{\underline{g}{}_s}
+ N \log \left[ \left( \frac{2\pi}{\underline{g}{}_s} \right)^{1/2}
\ex^{-1/\underline{g}{}_s} I_M(1/\underline{g}{}_s)\right] 
+\sum_{j=1}^N (N-j) \log \left( \frac{1 - R_j^2(M)}{j \underline{g}{}_s} \right).
\ee
Here we stress that \eqref{Fnpp} is an exact rearrangement of terms with no approximation.
Note that $\mathcal{F}^{(\mathrm{G})}$ is the free energy of the $N \times N$ Hermitian matrix model with Gaussian potential
\be
Z_G = \frac{1}{\mathrm{vol}(U(N))} \int [ \de H]
\exp\left( - \frac{1}{2 \,\underline{g}{}_s} \mathrm{Tr} H^2 \right) 
= \exp\bigl( \mathcal{F}^{(\mathrm{G})} \bigr),
\ee
and $\mathcal{F}^{(\mathrm{n})} = \mathcal{F}-\mathcal{F}^{(\mathrm{G})}$
is the ``normalized free energy'' (see, for example, \cite{mar05}).

Now let us consider the $\underline{g}{}_s$ perturbation of $\mathcal{F}^{(\mathrm{n})}$ 
ignoring the ``instanton contributions'', i.e., exponentially small terms. Let us denote the expansion
as follows:
\bel{NFE2}
\mathcal{F}^{(\mathrm{n})}
= \sum_{k=-1}^{\infty} 
\widetilde{\mathcal{F}}_k^{(\mathrm{n})}(N) \, \underline{g}{}_s{}^k 
+ O(\ex^{-2/\underline{g}{}_s}).
\ee
Obviously, we have
\be
\widetilde{\mathcal{F}}^{(\mathrm{n})}_{-1}(N) = N, \qq 
\widetilde{\mathcal{F}}^{(\mathrm{n})}_0(N) = 0.
\ee
We have determined $\widetilde{\mathcal{F}}^{(\mathrm{n})}_k$ up to $k=10$. 
The explicit form of $\widetilde{\mathcal{F}}^{(\mathrm{n})}_k$ are given by
\be
\widetilde{\mathcal{F}}^{(\mathrm{n})}_1 
= - \frac{1}{8}\bigl( (2M)^2-1 \bigr) N,
\ee
\be
\widetilde{\mathcal{F}}_2^{(\mathrm{n})} = -\frac{1}{16}\bigl( (2M)^2-1 \bigr) N^2,
\ee
\be
\widetilde{\mathcal{F}}^{(\mathrm{n})}_3
=-\frac{1}{24}\bigl( (2M)^2-1 \bigr) N^3 
 + \frac{1}{384} \bigl( (2M)^2-1 \bigr) \bigl( (2M)^2-3^2 \bigr)  N,
\ee
\be
\widetilde{\mathcal{F}}^{(\mathrm{n})}_4
=-\frac{1}{32}  \bigl( (2M)^2-1 \bigr) N^4
+ \frac{1}{128} \bigl( (2M)^2-1 \bigr)\bigl( (2M)^2-3^2 \bigr)N^2,
\ee
\be
\begin{split}
\widetilde{\mathcal{F}}^{(\mathrm{n})}_5
&= -\frac{1}{40} \bigl( (2M)^2-1 \bigr)  N^5 
+ \frac{1}{64}  \bigl( (2M)^2-1 \bigr) \bigl( (2M)^2-3^2 \bigr)  N^3 \cr
&- \frac{1}{5120} \bigl( (2M)^2-1 \bigr) \bigl( (2M)^2-3^2 \bigr) \bigl( (2M)^2-5^2 \bigr) N,
\end{split}
\ee
\be
\begin{split}
\widetilde{\mathcal{F}}^{(\mathrm{n})}_6
&= -\frac{1}{48}\bigl( (2M)^2-1 \bigr) N^6 
+ \frac{5}{192} \bigl( (2M)^2-1 \bigr)\bigl( (2M)^2-3^2 \bigr)N^4 \cr
&- \frac{1}{192} \bigl( (2M)^2-1 \bigr)\bigl( (2M)^2-3^2 \bigr)(M^2-6) N^2,
\end{split}
\ee
\be
\begin{split}
\widetilde{\mathcal{F}}^{(\mathrm{n})}_7
&= -\frac{1}{56} \bigl( (2M)^2-1 \bigr) N^7
+\frac{5}{128} \bigl( (2M)^2-1 \bigr)\bigl( (2M)^2-3^2 \bigr)N^5 \cr
&- \frac{1}{1024} \bigl( (2M)^2-1 \bigr)\bigl( (2M)^2-3^2 \bigr)(20\, M^2 - 117) N^3 \cr
&+ \frac{5}{229376} \bigl( (2M)^2-1 \bigr)\bigl( (2M)^2-3^2 \bigr)
\bigl( (2M)^2-5^2 \bigr)\bigl( (2M)^2-7^2 \bigr)N,
\end{split}
\ee
\be
\begin{split}
\widetilde{\mathcal{F}}^{(\mathrm{n})}_8
&= -\frac{1}{64}\bigl( (2M)^2-1 \bigr) N^8 
+ \frac{7}{128}\bigl( (2M)^2-1 \bigr)\bigl( (2M)^2-3^2 \bigr) N^6 \cr 
&- \frac{7}{512} \bigl( (2M)^2-1 \bigr)\bigl( (2M)^2-3^2 \bigr)(4\, M^2 - 23) N^4 \cr
& + \frac{1}{4096}
\bigl( (2M)^2-1\bigr)\bigl( (2M)^2-3^2 \bigr)\bigl( (2M)^2-5^2 \bigr)
(4\, M^2 -45) N^2,
\end{split}
\ee
\be
\begin{split}
\widetilde{\mathcal{F}}^{(\mathrm{n})}_9 
&=-\frac{1}{72}\bigl( (2M)^2-1 \bigr) N^9 
+ \frac{7}{96}\bigl( (2M)^2-1 \bigr)\bigl( (2M)^2-3^2 \bigr) N^7  \cr
&- \frac{7}{1536} \bigl( (2M)^2-1 \bigr)\bigl( (2M)^2-3^2 \bigr)(28\, M^2 - 159)
N^5 \cr
&+ \bigl( (2M)^2-1 \bigr)\bigl( (2M)^2-3^2 \bigr)
\left( \frac{35\, M^4}{1536} - \frac{3545\, M^2}{9216} 
+ \frac{111689}{73728} \right)
N^3 \cr
& - \frac{7}{2359296} \left[\prod_{k=0}^4 \bigl( (2M)^2 - (2k+1)^2 \bigr) \right] N,
\end{split}
\ee
\be
\begin{split}
\widetilde{\mathcal{F}}^{(\mathrm{n})}_{10} 
&=- \frac{1}{80} \bigl( (2M)^2-1 \bigr) N^{10}
+ \frac{3}{32}\bigl( (2M)^2-1 \bigr)\bigl( (2M)^2-3^2 \bigr) N^8 \cr
&- \frac{21}{640} \bigl( (2M)^2-1 \bigr)\bigl( (2M)^2-3^2 \bigr)(8\, M^2 - 45)
N^6 \cr
&+ \bigl( (2M)^2-1 \bigr)\bigl( (2M)^2-3^2 \bigr)
\left( \frac{3\, M^4}{32} - \frac{395\, M^2}{256} 
+ \frac{6133}{1024} \right)
N^4 \cr
&- \bigl( (2M)^2-1 \bigr)\bigl( (2M)^2-3^2 \bigr)\bigl( (2M)^2-5^2 \bigr)
\left( \frac{M^4}{1280} - \frac{3\, M^2}{128} + \frac{3497}{20480} \right) N^2.
\end{split}
\ee
Note that for $k \geq 1$, the leading and subleading terms are guessed as follows
\be
\begin{split}
\widetilde{\mathcal{F}}^{(\mathrm{n})}_k &= -\frac{1}{8k} \bigl( (2M)^2-1 \bigr)  N^k  \cr
&+ \frac{(k-1)(k-2)}{768} \bigl( (2M)^2-1 \bigr)\bigl( (2M)^2-3^2 \bigr)
N^{k-2} + \dotsm.
\end{split}
\ee

Let $\widetilde{S}:= N \underline{g}{}_s$. The normalized free energy \eqref{NFE2}
can be rearranged as follows:
\be
\mathcal{F}^{(\mathrm{n})}
= \sum_{g=0}^{\infty} \mathcal{F}^{(\mathrm{n})}_g(\widetilde{S}) \, 
\underline{g}{}_s{}^{2g-2} +O(\ex^{-2/\underline{g}{}_s}).
\ee
Then, we have determined $\mathcal{F}^{(\mathrm{n})}_g$ up to $g=3$ by extrapolating our perturbative
calculation:
\be
\begin{split}
\mathcal{F}^{(\mathrm{n})}_0(\widetilde{S})&= \widetilde{S}, \cr
\mathcal{F}^{(\mathrm{n})}_1(\widetilde{S})&=
\frac{1}{8}\bigl( (2M)^2-1 \bigr) \log(1 - \widetilde{S} ), \cr
\mathcal{F}^{(\mathrm{n})}_2(\widetilde{S})&= 
\frac{1}{384} \bigl( (2M)^2-1 \bigr) \bigl( (2M)^2-3^2 \bigr) 
\frac{\widetilde{S}}{(1- \widetilde{S})^3}, \cr
\mathcal{F}^{(\mathrm{n})}_3(\widetilde{S})&=
- \frac{\bigl( (2M)^2-1 \bigr) \bigl( (2M)^2-3^2 \bigr)}{15360}
\Bigl[3 \bigl((2M)^2-5^2 \bigr)
+ 2 \bigl(4\, M^2-15 \bigr) \widetilde{S} \Bigr] \frac{\widetilde{S}}{(1-\widetilde{S})^6}.
\end{split}
\ee
When $M=0$, these results coincide with the free energy of the GWW model in the weak coupling
phase ($\widetilde{S}<1$) \cite{gol80,mar0805}.

We expect that for $g \geq 2$, the $\mathcal{F}_g^{(\mathrm{n})}$ takes the
following form:
\be
\mathcal{F}_g^{(\mathrm{n})}
= \frac{(\mbox{a polynomial of $\widetilde{S}$})}{(1 - \widetilde{S})^{3(g-1)}},
\ee
where $3(g-1) = (2 - \gamma_{\mathrm{st}})(g-1)$ with $\gamma_{\mathrm{st}}=-1$
is the susceptibility of this model.

We remark that by setting $M$ to a half-integer, the expression of the free energy takes a simpler form.
In particular at
$M = \pm 1/2$, it seems that $\mathcal{F}^{(\mathrm{n})}=\widetilde{S} \, \underline{g}{}_s^{-2}$.
Hence, it may be useful to rewrite the $M$-dependence of $\mathcal{F}^{(\mathrm{n})}$ 
similar to \eqref{RnM2}:
\be
\mathcal{F}^{(\mathrm{n})} = \frac{1}{\pi} \cos ( \pi M)
\left( \sum_{j=0}^{\infty} \frac{(-1)^j(2j+1)}{\{(j+1/2)^2 -M^2\}}
\mathcal{F}^{(\mathrm{n})} \Bigr|_{M=j+1/2} \right).
\ee

\section{Various limits of $N_f=2$ matrix model}

\subsection{Planar limit of \eqref{STREQ} and critical points}

The planar limit is taken by sending $N \rightarrow \infty$, $\underline{g}{}_s \rightarrow 0$
with $\widetilde{S}=N \tilde{g}{}_s$ kept finite.
In the planar limit,  $\eta_n=n \, \tilde{g}{}_s$ turns into a continuous function of $x\equiv n/N$:
$\eta_n \rightarrow \eta(x) = \widetilde{S}\, x$ with $0 \leq x \leq 1$.  We assume that $\xi_n$ also becomes a
continuous function of $x$:
\be
\xi_n=\xi\left( \frac{n}{N} \right) \rightarrow \xi(x).
\ee

In the planar limit, the second line of \eqref{STREQ} is ignored because
\bel{xix}
\xi_{n \pm 1} = \xi\left( x \pm \frac{1}{N} \right) \rightarrow \xi(x).
\ee
Hence the following quartic equation for $\xi$ is obtained from \eqref{STREQ}:
\bel{xiq}
\xi^2 ( 1 - \xi)^2 - \eta^2\, \xi^2 + \zeta^2 (1  - \xi)^2 = 0.
\ee
The discriminant of this equation with respect to $\xi$ is given by
\be
\Delta = 16\, \eta^2 \, \zeta^2
\Bigl( \eta^2 - (1 + \zeta^{2/3})^3 \Bigr)
\Bigl( \eta^2 - (1 + \omega \, \zeta^{2/3})^3 \Bigr)
\Bigl( \eta^2 - (1 + \omega^2 \, \zeta^{2/3})^3 \Bigr),
\ee
where $\omega \equiv \exp(2\pi \im/3)$. At $\eta=\pm 1$, $\zeta=0$,
the quartic equation \eqref{xiq} turns into
\be
\xi^3(\xi-2) = 0,
\ee
and the three roots out of four degenerate to $\xi=0$. Therefore, we choose the
critical values of $(\xi, \eta, \zeta)$ as
\bel{critxez}
\xi_{\mathrm{c}} = 0, \qq
\eta_{\mathrm{c}} = \pm 1, \qq
\zeta_{\mathrm{c}} = 0,
\ee
where we take the continuum limit.

\subsubsection{Planar free energy and susceptibility}

Using \eqref{xix}, the planar free energy can be written as
\bel{PFE}
\begin{split}
F &\equiv \lim_{N \rightarrow 0} \left( -\frac{1}{N^2} \log\Bigl( (-1)^{MN}  \underline{Z}{}_{U(N)} \Bigr) 
\right) \cr
&= - \lim_{N \rightarrow \infty}  \frac{1}{N} \log \Bigl( I_M(1/\underline{g}{}_s) \Bigr)
- \lim_{N \rightarrow \infty}\sum_{j=1}^N
\frac{1}{N} \left( 1 - \frac{n}{N} \right) \log (1 - \xi_n ) \cr
&=- \frac{1}{\widetilde{S}} - \int_0^1 (1-x) \log\bigl( 1 - \xi(x) \bigr) \, \de x.
\end{split}
\ee
Note that this planar free energy is related to the planar part of the free energy $\mathcal{F}$ \eqref{UFE}
as
 $\mathcal{F}_0 = -\widetilde{S}^2 \, F$, where 
\bel{PFE2}
\mathcal{F}= \sum_{g=0}^{\infty} \mathcal{F}_g(\widetilde{S}) \underline{g}{}_s{}^{2g-2}, \qq
\mathcal{F}_g = \mathcal{F}_g^{(\mathrm{G})} + \mathcal{F}_g^{(\mathrm{n})}.
\ee
Let  $a \equiv N^{-1/3}$.
Setting $\xi = a^2 \, U$, $\eta = \pm (1 -a^2 t)$, $\zeta = a^3 z$,
\eqref{STREQ} leads to the following equation at $\mathcal{O}(a^6)$:
\begin{align} \label{pleq}
	t = U - \frac{z^2}{2\, U^2}.
\end{align}
For simplicity, we consider the case $\eta = 1 - a^2 t = \widetilde{S}\, x$.
Then, for small $a>0$, \eqref{pleq} is equivalent to
\be
1 - \widetilde{S}\, x = \xi - \frac{a^6\, z^2}{2\, \xi^2}.
\ee
This leads to a cubic equation for $\xi$.
Let $\xi(x; a)$ be one of three solutions to the cubic equation.
Note that
\be
\xi(x; 0) = \lim_{a \rightarrow 0} \xi(x; a) = 1 - \widetilde{S}\, x.
\ee
Hence, in the planar limit, we have
\be
-\lim_{a \rightarrow 0}  \int_0^1 (1-x) \log\bigl( 1- \xi(x; a) \bigr) \, \de
x = - \int_0^1 (1-x) \log ( \widetilde{S} x )
= \frac{3}{4} - \frac{1}{2} \log \widetilde{S}. 
\ee
Therefore, the planar free energy \eqref{PFE} is evaluated as
\be
-F = \frac{1}{\tilde{S}} + \frac{1}{2} \log \widetilde{S} - \frac{3}{4}.
\ee
Equivalently, the planar part of the free energy \eqref{PFE2} is equal to
\bel{F0p}
\mathcal{F}_0(\widetilde{S}) = - \widetilde{S}^2\, F =  \widetilde{S} + \widetilde{S}^2
\left( \frac{1}{2} \log \widetilde{S} - \frac{3}{4} \right).
\ee
Recall that the planar free energy
of the GWW model is given by
\be
\mathcal{F}_0(\widetilde{S}) 
=\begin{cases}
\dis 1/4, & ( \widetilde{S} \geq 1), \cr
\dis \widetilde{S} + (1/2) \widetilde{S}^2\left( \log \widetilde{S} - (3/2) \right),
& ( \widetilde{S} \leq 1).
\end{cases}
\ee
Hence the planar free energy \eqref{F0p} coincides with
that of GWW model in the weak coupling phase $ (\widetilde{S} \leq 1)$ and is independent on $z$.
(The parameter $z$ is proportional to $M$).

In the planar limit, we have $\xi = 1 - \widetilde{S}\, x$, hence
\be
\begin{split}
&-\int_0^1 (1-x) \log(1-\xi(x) ) \, \de x \cr 
&=- \frac{1}{\widetilde{S}^2}
\int_{1- \widetilde{S}}^1 \bigl[ \xi - (1-\widetilde{S}) \bigr] \log (1 - \xi )\, \de \xi \cr
&= \frac{1}{\widetilde{S}^2} \Bigl[ - \bigl\{ G_1(1) - (1 - \widetilde{S}) G_0(1) \bigr\}
+ \bigl\{ G_1(1-\widetilde{S}) - (1- \widetilde{S}) G_0(1- \widetilde{S}) \bigr\} \Bigr],
\end{split}
\ee
where
\be
G_1(\xi):=\int \xi \log(1-\xi) \, \de \xi 
=\frac{1}{2} ( 1- \xi)^2 \log(1-\xi)
- (1-\xi) \log(1-\xi) - \frac{1}{4} \xi^2 - \frac{1}{2} \xi,
\ee
\be
G_0(\xi):=\int \log(1-\xi)\, \de \xi
= - (1- \xi) \log(1-\xi) - \xi.
\ee
Here we have fixed the ambiguity in the integration constants by taking them to be zero.
Note that
\be
G_1(\xi) = - \frac{1}{3}\, \xi^3 + O(\xi^4), \qq
G_0(\xi) = -\frac{1}{2} \xi^2 - \frac{1}{6} \xi^3 + O(\xi^4).
\ee
Then, the contribution to the free energy around $x=1$ (i.e., around $\xi= 1- \widetilde{S}$)
is
\bel{FLTx1}
\begin{split}
&\frac{1}{\widetilde{S}^2}\bigl\{ G_1(1-\widetilde{S}) - (1- \widetilde{S}) G_0(1- \widetilde{S}) \bigr\} \cr
&=\frac{1}{\widetilde{S}^2}
\left( -\frac{1}{3} (1 - \widetilde{S})^3 + \frac{1}{2} (1 - \widetilde{S})^3 +O\bigl((1-\widetilde{S})^4 \bigr)
\right) \cr
&= \frac{1}{6}  (1 - \widetilde{S})^3 + O\bigl((1-\widetilde{S})^4 \bigr).
\end{split}
\ee
Thinking  the critical value of the $\widetilde{S}$ as $\widetilde{S}_{\mathrm{c}}=1$,
we can read off the susceptibility  $\gamma_{\mathrm{st}}$ from the leading term (see, for example, \cite{LAG,GM9304}):
\be
( 1- \widetilde{S})^3 = (\widetilde{S}_{\mathrm{c}} - \widetilde{S})^{2 - \gamma_{\mathrm{st}}},
\qq \gamma_{\mathrm{st}} = - 1.
\ee
The susceptibility $\gamma_{\mathrm{st}}$ is $-1$ and independent of $z$.

Let us denote the planar free energy of GWW model in the strong and weak coupling phases respectively by
\be
F_{\mathrm{s}} = - \frac{1}{4 \widetilde{S}^2}, \qq
(\widetilde{S} \geq 1),
\ee
\be
F_{\mathrm{w}}
= \frac{3}{4} - \frac{1}{\widetilde{S}} - \frac{1}{2} \log \widetilde{S}, \qq
( \widetilde{S} \leq 1).
\ee
Note that
\be
\begin{split}
F_{\mathrm{s}}
&= -\frac{1}{4} - \frac{1}{2} (1- \widetilde{S})
- \frac{3}{4} (1 - \widetilde{S})^2 - (1- \widetilde{S})^3
+ O\bigl( (1 - \widetilde{S})^4 \bigr), \cr
F_{\mathrm{w}}
&= -\frac{1}{4} - \frac{1}{2} (1- \widetilde{S})
- \frac{3}{4} (1 - \widetilde{S})^2 - \frac{5}{6} (1- \widetilde{S})^3
+ O\bigl( (1 - \widetilde{S})^4 \bigr).
\end{split}
\ee 
So, at least in the leading order in small $(1- \widetilde{S})$,
\eqref{FLTx1} is equal to the discontinuity of the planar free energy:
\be
F_{\mathrm{w}}  - F_{\mathrm{s}}
= \frac{1}{6}  (1 - \widetilde{S})^3 + O\bigl((1-\widetilde{S})^4 \bigr).
\ee

\subsubsection{Singular K3 surface}

In the planar limit, eq.\eqref{STREQ} also becomes the defining relation of an algebraic variety.
With the introduction of the homogeneous coordinates $(\xi:\eta:\zeta:1)=(\mathcal{X}:\mathcal{Y}:\mathcal{Z}:\mathcal{W})$ of 
the three-dimensional complex projective space $\mathbb{P}^3$,
this algebraic variety is the union of the hyperplane $\mathcal{Y}=0$ (with multiplicity two) and the singular K3 surface
\be
- \mathcal{Y}^2\, \mathcal{X}^2  +\mathcal{X}^2 (\mathcal{X}-\mathcal{W})^2 + (\mathcal{X}-\mathcal{W})^2 \mathcal{Z}^2= 0.
\ee
The singular loci of this surface consist of three spheres:
\be
\begin{split}
S^2_{(1)} &= \{ (0: b: 0: d) \in \mathbb{P}^3 \,  |\, (b:d) \in \mathbb{P}^1 \}, \cr
S^2_{(2)} &= \{ (0: b: c: 0) \in \mathbb{P}^3 \, |\, (b:c) \in \mathbb{P}^1 \}, \cr
S^2_{(3)} &= \{ (a: 0: c: a) \in \mathbb{P}^3 \, |\, (a:c) \in \mathbb{P}^1 \}.
\end{split}
\ee
The intersections of these spheres are given by
\be
\begin{split}
S^2_{(1)} \cap S^2_{(2)} &= \{ (0:1:0:0) \}, \cr
S^2_{(1)} \cap S^2_{(3)} &= \emptyset, \cr
S^2_{(2)} \cap S^2_{(3)} &= \{ (0:0:1:0) \}.
\end{split}
\ee
The appearance of the singular K3 surface is just an observation. At this moment,
we have no idea about
its geometrical or physical meaning. 
The alternate discrete Painlev\'{e} II equation is closely related to a rational surface,
called the space of initial values \cite{oka79,sak01}, characterized by the affine root system $D^{(1)}_6$.
The relation between these two surfaces is still not known to us.

We remark that the critical point \eqref{critxez} is on the first sphere $S^2_{(1)}$:
\be
(\xi_{\mathrm{c}}: \eta_{\mathrm{c}}: \zeta_{\mathrm{c}}: 1)
= ( 0: \pm 1: 0: 1 ) \in S^2_{(1)}.
\ee

\subsection{Double scaling limit}

Note that
\be
\widetilde{S} = N \underline{g}{}_s = \frac{g_s \, N }{\Lambda_2} = - \frac{(m_1 + m_2) }{\Lambda_2}.
\ee
This is the parameter we fine tune to $\pm 1$,
and is the counterpart of the bare cosmological constant in $2$d gravity.
Also note that 
\be
\zeta = M \underline{g}{}_s= \frac{M g_s}{\Lambda_2}=
\frac{(m_2 - m_1)}{\Lambda_2} = \mathcal{O}(a^3)
\ee
and the two masses are fine tuned to be equal in this limit.

\subsubsection{Double scaling limit and Painlev\'{e} II equation}

Let us consider the double scaling limit of \eqref{STREQ}. Let $x \equiv n/N$, 
$a^3 \equiv 1/N$ and
\be
\eta_n = \widetilde{S}\, x=1 - (1/2) a^2 \, t, \qq \zeta = a^3\, \widetilde{S}\, M, \label{DLS1}
\ee
\be
\xi(x) = \xi(n/N) =\xi_n = a^2 \, u(t).
\ee

With these scaling ansatz, 
the double scaling limit is defined as the $N \rightarrow \infty$ ($a \rightarrow 0$)  limit
while simultaneously sending $\widetilde{S}$ to its critical value $1$ by \eqref{DLS1}.
The original 't Hooft expansion parameter $1/N$ gets dressed by the combination which is kept finite in this limit:
\begin{align}
	\kappa \equiv \frac{1}{N} \frac{1}{(1 - \widetilde{S})^{(1/2)(2 - \gamma_{\mathrm{st}})}},
	\qq \gamma_{\mathrm{st}} = -1
\end{align}
with $\gamma_{\mathrm{st}}$ being the susceptibility of the system. For later convenience, we write
\bel{tilc}
\widetilde{S}=1 - \tilde{c}\, a^2, \qq
\kappa=\tilde{c}{}^{-3/2}.
\ee
Note that
\be
\xi_{n \pm 1} = \xi((n\pm 1)/N) = \xi(x \pm (1/N) )
= \xi(x) \pm \frac{1}{N} \frac{\partial \xi(x)}{\partial x} + 
\frac{1}{2N^2} \frac{\partial^2 \xi(x)}{\partial x^2}+
O(1/N^3) ,
\ee
and
\be
\frac{\partial t}{\partial x} = - \frac{2 \, \widetilde{S}}{a^2}.
\ee
We can see that
\be
\frac{1}{N} \frac{\partial \xi(x)}{\partial x}
= a^3 \cdot \frac{\partial t}{\partial x} \frac{\partial (a^2 u(t))}{\partial t}
= - 2 \, \widetilde{S} a^3 u'(t),
\ee
\be
\frac{1}{2N^2} \frac{\partial^2 \xi(x)}{\partial x^2}
= \frac{a^6}{2} \left( \frac{\partial t}{\partial x} \frac{\partial}{\partial t}
\right)^2 a^2 u(t)
= 2 \, \widetilde{S}^2 \, a^4 u''(t).
\ee 
Hence, the double scaling limit of $\xi_{n \pm 1}$ is given by
\bel{xipm}
\xi_{n \pm 1} \rightarrow a^2 u(t) \mp 2 \, a^3 u'(t)  + 2 \, a^4 u''(t) + O(a^5).
\ee
Using $\eta_n \rightarrow 1-(1/2) a^2 t$, $\zeta \rightarrow a^3 M$, $\xi_n \rightarrow a^2 u(t)$
and \eqref{xipm}, the string equation \eqref{STREQ} turns into
\be
0 = \Bigl( t u^2 - 2 u^3 + M^2 + 2 \, u u'' - (u')^2 \Bigr) a^6 + O(a^8).
\ee
Note that the value of $\tilde{c}$, introduced in \eqref{tilc}, is irrelevant to this order.

Therefore, we obtain the Painlev\'{e} II equation as the double scaling limit of \eqref{STREQ}:
\bel{PIIu}
u'' = \frac{(u')^2}{2\, u} + u^2 - \frac{1}{2} \, t\, u - \frac{M^2}{2\, u}.
\ee
This is the PII equation obtained from the Lax pair of Flaschka-Newell \cite{FN80}.
It can be converted into more standard form of PII \eqref{SPII} as follows.
By using $p_u \equiv -u'/u$, eq. \eqref{PIIu} is equivalent to the following system of equations
\be
\begin{split}
u' &= - p_u u, \cr
p_u' &= \frac{1}{2} \, p_u^2 - u + \frac{1}{2} \, t + \frac{M^2}{2u^2}.
\end{split}
\ee
This is
a Hamilton system 
\be
u' = \frac{\partial H_{\mathrm{II}}}{\partial p_u}, \qq
p_u' = - \frac{\partial H_{\mathrm{II}}}{\partial u},
\ee
with
the Hamiltonian
\bel{ham}
H_{\mathrm{II}}(u,p_u, t) = - \frac{1}{2}\, p_u^2 \, u + \frac{1}{2}\, u^2 - \frac{1}{2}\, t\, u
+ \frac{M^2}{2\, u}.
\ee
By the canonical transformation $(u,p_u) \rightarrow (y, p_y)$ with $u=-p_y$ and 
$p_u = y + (M/p_y$), this Hamiltonian becomes the following form \cite{mal22,oka86}
\be
H_{\mathrm{II}}= \frac{1}{2} \, p_y^2 + \frac{1}{2}\bigl( y^2 + t \bigr) p_y + M\, y,
\ee
and the system of equations is converted into 
\be
\begin{split}
y'&= p_y + \frac{1}{2} \, y^2 + \frac{1}{2} \, t, \cr
p_y' &= - p_y\, y - M.
\end{split}
\ee 
This means that $y=y(t)$ obeys the following form of the Painlev\'{e} II equation:
\bel{PIIs}
y'' = \frac{1}{2} \, y^3 + \frac{1}{2} \, t\, y + \left( \frac{1}{2} - M \right).
\ee
By appropriate rescaling of $y$ and $t$, this can be converted into the standard form of PII:
\bel{SPII}
y'' = 2 \, y^3 + t \, y + \alpha,
\ee
with $\alpha = (1/2) - M$.

When there is no logarithmic potential ($M=0$), 
the appearance of the Painlev\'{e} II equation in the unitary matrix model
was shown in \cite{PS90a,PS90b}. (See also, \cite{CKV0508}.)
In this case, by setting $u=y^2$, \eqref{PIIu} turns into
\be
y'' = \frac{1}{2} y^3 - \frac{t}{4} y.
\ee
Hence in \cite{PS90a,PS90b}, this is interpreted as the zero parameter ($\alpha=0$) case of PII.
But it can be also interpreted as the $\alpha=1/2$ case of PII.

\subsubsection{Double scaling limit and Argyres-Douglas point}

It is easy to see what this critical point corresponds to in the Seiberg-Witten curve (quartic one), 
which is the spectral curve obtained the planar loop equation/Virasoro constraints.
For the $N_f=2$ potential $W^{(2)}(z)$ \eqref{potNf2}, let
\be
W(z) := \lim_{g_s \rightarrow 0} g_s \, W^{(2)}(z)
= - \frac{\Lambda_2}{2} \left( z + \frac{1}{z} \right)+ 2 \, m_2 \log z.
\ee 
The spectral curve \eqref{PlanarCurve} for this potential $W(z)$ is given by
\bel{Nf2SWG1}
	y^2 = \frac{\Lambda_2{}^2}{16z^4} \left( 1 + \frac{8 \, m_2}{\Lambda_2} z 
	+ \frac{16\, \mathfrak{u}}{\Lambda_2{}^2} z^2 + \frac{8 \, m_1}{\Lambda_2} z^3 + z^4 \right).
\ee
This is the ``first realization'' of the Seiberg-Witten curve in the Gaiotto form for the $N_f=2$ theory \cite[eq.(10.11)]{GMN0907}.
The Seiberg-Witten differential is given by $d S_{\mathrm{SW}}=y(z)\, \de z$.
The coefficient of $z^{-2}$ in the right-handed side is identified
with the Coulomb moduli parameter $\mathfrak{u} = \langle \mathrm{Tr} \, \phi^2 \rangle$
of the Seiberg-Witten curve.
The parameter $\mathfrak{u}$ is related to the (planar) resolvent $\omega(z)$ \eqref{presol} as follows:
\be
\omega(z) = - \frac{(m_1+m_2)}{z} + \frac{\omega_1}{z^2} + O(z^{-3}),
\ee
where 
\be
\omega_1= \lim_{g_s \rightarrow 0} 
\dket g_s \sum_{I=1}^N w_I \dbr
= \frac{2(m_1^2-\mathfrak{u})}{\Lambda_2} - \frac{1}{4} \Lambda_2.
\ee

Clearly, at our critical point $m_1 / \Lambda_2 = m_2 / \Lambda_2 = \mp 1/2$, 
this genus one curve \eqref{Nf2SWG1} shrinks to a point at $\mathfrak{u} / \Lambda_2{}^2= 3/8$:
\be
y^2 = \frac{\Lambda_2{}^2}{16\, z^4} (z \mp 1)^4.
\ee
For simplicity, 
let us consider the limit to the following critical values:
\be
m_1=m_2 = - (1/2) \Lambda_2, \qq
\mathfrak{u}=(3/8) \Lambda_2{}^2, \qq
 \widetilde{S}=1, \qq 
 z = 1, \qq y=0.
\ee
Recall that
\be
\widetilde{S} = - \frac{(m_1+m_2)}{\Lambda_2}= 1 - \tilde{c}\, a^2,
\qq
\zeta= \frac{(m_2-m_1)}{\Lambda_2} = \widetilde{S} M\, a^3.
\ee
In the double scaling limit, these mass parameters are given by
\be
m_1 = - \frac{1}{2}\Lambda_2 (1 - \tilde{c} \, a^2) (1 + M\, a^3),
\qq
m_2 = - \frac{1}{2} \Lambda_2 (1 - \tilde{c}\, a^2) (1 - M\, a^3).
\ee

We require that $z$ and $\mathfrak{u}$ approach their critical values respectively by
\be
z = 1 -2\, a\, \tilde{z},
\qq
\mathfrak{u} = \Lambda_2{}^2 \left( \frac{3}{8} - \frac{1}{2} \, a^2\, \tilde{c}
+ a^4\, \tilde{u} \right).
\ee
Then we can see that
\be
\frac{\Lambda_2{}^2}{16 z^4}\left(
1 + \frac{8\, m_2}{\Lambda_2}z + \frac{16 \, \mathfrak{u}}{\Lambda_2{}^2} z^2
+ \frac{8 \, m_1}{\Lambda_2} z^3 + z^4 \right)
= \Lambda_2{}^2 \Bigl( \tilde{z}^4 + \tilde{c}\, \tilde{z}^2
+M\, \tilde{z}+ \tilde{u} \Bigr) a^4 +O(a^5).
\ee
Hence by setting
\be
y =\Lambda_2\, \tilde{y}\, a^2,
\ee
the curve \eqref{Nf2SWG1} in $a \rightarrow 0$ limit turns into
the first realization of 
the Seiberg-Witten curve for the Argyres-Douglas theory of type $H_1=(A_1, A_3)$ \cite{APSW}:
\be
\tilde{y}^2 = \tilde{z}^4 + \tilde{c}\, \tilde{z}^2
+M\, \tilde{z}+ \tilde{u}.
\ee
Our limit is, therefore, the limit toward the $H_1$ Argyres-Douglas point .

\section*{Acknowledgments}

We thank Chuan-Tsung Chan, Masafumi Fukuma, Hikaru Kawai, Vladimir Kazakov, Kazunobu Maruyoshi, Shun'ya Mizoguchi, Hajime Nagoya and Kentaroh Yoshida for useful discussions.
We also thank the anonymous referee of our previous paper \cite{IOYanok1} for valuable suggestions.


\appendix

\section{The volume of the unitary group}
\label{VUN}

In this section, we show that the volume of the $U(N)$ group, constructed from
the measure \eqref{UNmet}, is given by \eqref{volUN}.

A unitary matrix $U \in U(N)$ can be expressed by an $N \times N$ Hermitian matrix $H$ as follows \cite{MMS91,miz0411}
\bel{UtoH}
U = \frac{1 + \im \, H}{1 - \im \, H}, \qq
U^{-1} = U^{\dag} = \frac{1 - \im \, H}{1 + \im \, H}.
\ee
Let $[\de U]$ be a Haar measure for the unitary matrix defined by the metric
\bel{dsU}
\de s^2 = \mathrm{Tr}\bigl( \de U^{\dag}\, \de U \bigr) = - \mathrm{Tr}\bigl( U^{-1} \de U\bigr)^2.
\ee 
Also let $[\de H]$ be a measure for the Hermitian matrix defined by the metric $\mathrm{Tr}\, ( \de H)^2$.
Then, \eqref{UtoH} leads to the following relation between the two measures:
\be
[ \de U ] = \frac{2^{N^2}}{\dis \Bigl( \det(1 + H^2) \Bigr)^N} [ \de H].
\ee
The unitary matrix $U$ and the Hermitian matrix $H$ can be diagonalized by a unitary matrix $V$:
\begin{align}
U &= V^{-1} U_D V,&
U_D &= \mathrm{diag}(\ex^{\im\,  \theta_1}, \ex^{\im \, \theta_2}, \dotsm, \ex^{\im \, \theta_N}), &
( -\pi \leq &\theta_i \leq \pi), \cr
H &= V^{-1} \Lambda V, &
\Lambda &= \mathrm{diag}(\lambda_1, \lambda_2 ,\dotsm, \lambda_N), &
( -\infty < &\lambda_i < \infty), 
\end{align}
and the eigenvalues are related as
\bel{evUH}
\ex^{\im \, \theta_i} = \frac{1 + \im \, \lambda_i}{1- \im \, \lambda_i}, \qq
\lambda_i = \tan \frac{\theta_i}{2}.
\ee
Note that there is an ambiguity in the choice of $V$. For a diagonal matrix $A_D \in U(N)$, $A_D V$ also
diagonalizes $U$ and $H$.

Let us write the measure for the Hermitian matrix as
\be
[ \de H ] = \Delta(\lambda)^2 \prod_{i=1}^N \de \lambda_i \times [ \de \Omega_N], \qq
\Delta(\lambda) = \prod_{1\leq i<j \leq N} ( \lambda_i - \lambda_j).
\ee
Using the Gaussian integral
\be
\int [ \de H] \, \exp\left( -\frac{1}{2\, g_s}\mathrm{Tr} \, H^2 \right)=(2 \, \pi \, g_s)^{(1/2)N^2},
\ee
and an integral formula
\be
\int \de^{N} \lambda \, |\Delta(\lambda)|^{2 \beta} \exp\left( - \frac{1}{2} \sum_i \lambda_i^2 \right)
= (2\pi)^{N/2} \prod_{j=1}^N \frac{\Gamma(\beta  j+1)}{\Gamma(\beta+1)},
\ee
we can show that
\be
\Omega_N:= \int [ \de \Omega_N ] = \frac{(2\pi)^{(1/2)N(N-1)}}{G_2(N+2)}.
\ee
Here $G_2(z)$ is the Barnes function, defined by
$G_2(z+1) = \Gamma(z) G_2(z)$ with $G_2(1)=1$.

Using another integral formula
\be
\int \prod_{i=1}^N \frac{\de \lambda_i}{(1+\lambda_i^2)^N} \, \Delta(\lambda)^2
=\frac{N!}{2^{N(N-1)}} \pi^N,
\ee
we can evaluate the volume of the unitary group $U(N)$:
\be
\begin{split}
\mathrm{vol}(U(N)) &= \int [ \de U] 
= 2^{N^2} \int [ \de H] \Bigl( \det(1+H^2) \Bigr)^{-N} \cr
&= 2^{N^2}\, \int \prod_{i=1}^N \frac{\de \lambda_i}{(1+\lambda_i^2)^N}\,  \Delta(\lambda)^2 
[ \de \Omega_N] 
=2^{N^2} \times \frac{N!}{2^{N(N-1)}} \pi^N \times \Omega_N.
\end{split}
\ee
Hence the volume of the unitary group obtained from the metric \eqref{dsU} is given by
\be
\mathrm{vol}(U(N)) = (2\pi)^N \, N!\, \Omega_N = \frac{(2\pi)^{(1/2)N(N+1)}}{G_2(N+1)}.
\ee
This completes the proof of \eqref{volUN}. This volume
coincides with the one adopted in the $U(N)$ Chern-Simons theory \cite{OV0205}.

Using \eqref{evUH}, we can see that
\be
\begin{split}
2^{N^2} \, \prod_{i=1}^N \frac{\de \lambda_i}{(1+\lambda_i^2)^N} \prod_{1 \leq i<j \leq N}
(\lambda_i - \lambda_j)^2 
&= 2^{N(N-1)} \prod_{i=1}^N \de \theta_i \prod_{1 \leq i<j \leq N}
\sin^2 \left( \frac{\theta_i - \theta_j}{2} \right) \cr
&=\prod_{i=1}^N \de \theta_i\, \Delta(\theta) \overline{\Delta(\theta)},
\end{split}
\ee
where
\be
\Delta(\theta) = \prod_{1 \leq i< j \leq N} \bigl( \ex^{\im \theta_i} - \ex^{\im \theta_j} \bigr),
\qq
\overline{\Delta(\theta)} = \prod_{1 \leq i< j \leq N} \bigl( \ex^{-\im \theta_i} - \ex^{-\im \theta_j} \bigr).
\ee
Therefore, the unitary Haar measure can be written as
\be
[ \de U] = \Delta(\theta)\,  \overline{\Delta(\theta)} \left( \prod_{i=1}^N \de \theta_i \right) 
[ \de \Omega_N ].
\ee
It holds that
\be
\int \de^N \theta \, \Delta(\theta)\, \overline{\Delta(\theta)} = (2\pi)^N\, N!.
\ee

\section{Derivation of string equations}

\subsection{Derivation of \eqref{strcnstm1s}, \eqref{strcnst0s} and \eqref{strcnst1s}}
\label{DSE1}

The left-handed side of \eqref{strcnstm1s} can be written as
\bel{Beq1}
\int \de \mu(z) \left( z \frac{\partial}{\partial z} p_n(z) \right) z^{-1} \tilde{p}_{n-1}(1/z)
+ \int \de \mu(z) p_n(z) \left( z \frac{\partial}{\partial z} ( z^{-1} \tilde{p}_{n-1}(1/z) ) \right). 
\ee
Note that
\be
\begin{split}
z \frac{\partial}{\partial z} p_n(z) &= z \frac{\partial}{\partial z} \Bigl( z^{n} +( \text{lower power terms in $z$}) \Bigr) \cr
&=n z^{n} + \text{(lower power terms in $z$)}.
\end{split}
\ee
Thus the first term of \eqref{Beq1} is evaluated as
\bel{B1a}
\begin{split}
& \int \de \mu(z) \left( z \frac{\partial}{\partial z} p_n(z) \right) z^{-1} \tilde{p}_{n-1}(1/z) \cr
&= n \int \de \mu(z) z^{n-1} \tilde{p}_{n-1}(1/z) 
= n \int \de \mu(z) p_{n-1}(z) \tilde{p}_{n-1} = n  h_{n-1}.
\end{split}
\ee
Here we have used \eqref{pp1}.

Also, by using
\be
\begin{split}
z \frac{\partial}{\partial z} ( z^{-1} \tilde{p}_{n-1}(1/z) )
&= z \frac{\partial}{\partial z} (z^{-n} + \text{(lower power terms in $z^{-1}$)} \cr
&= -n z^{-n} +  \text{(lower power terms in $z^{-1}$)},
\end{split}
\ee
the second term of \eqref{Beq1} is given by
\bel{B1b}
\int \de \mu(z) p_n(z) \left( z \frac{\partial}{\partial z} ( z^{-1} \tilde{p}_{n-1}(1/z) ) \right)
=-n \int \de \mu(z) p_n(z) z^{-n} = - n h_{n}.
\ee
By adding \eqref{B1a} and \eqref{B1b}, we obtain \eqref{strcnstm1s}.

The remaining equations \eqref{strcnst0s} and \eqref{strcnst1s} can be obtained similarly.

\subsection{Derivation of \eqref{USE1}, \eqref{USE2} and \eqref{USE3}}
\label{DSE2}

Eq. \eqref{USE1} can be obtained by using the following relations:
\be
\int \de \mu(z) p_n(z) \tilde{p}_{n-1}(1/z) = 0,
\ee
\be
\int \de \mu(z) p_n(z) z^{-1} \tilde{p}_{n-1}(1/z)
=\int \de \mu(z) p_n(z) \left( \tilde{p}_n(1/z) + \sum_{k=0}^{n-1} \widetilde{C}^{(n-1)}_k \tilde{p}_k(1/z) \right)
= h_n,
\ee
\be
\begin{split}
& \int \de \mu(z) p_n(z) z^{-2} \tilde{p}_{n-1}(1/z) \cr
&=\int \de \mu(z) p_n(z) z^{-1} \left( \tilde{p}_n(1/z) + \widetilde{C}^{(n-1)}_{n-1} \tilde{p}_{n-1}(1/z)
+ \dotsm  \right) \cr
&=\int \de \mu(z) p_n(z) \left( \tilde{p}_{n+1}(1/z) + \widetilde{C}^{(n)}_n \tilde{p}_{n}(1/z)
+ \widetilde{C}^{(n-1)}_{n-1} \tilde{p}_n(1/z) + \dotsm \right) \cr
&= \Bigl( \widetilde{C}^{(n)}_n + \widetilde{C}^{(n-1)}_{n-1} \Bigr) h_n.
\end{split}
\ee
Here we have used the orthogonality \eqref{OC} and \eqref{recp2a}.

The remaining equations \eqref{USE2} and \eqref{USE3} can be derived similarly.

\section{alt-dPII}
\label{altdPII}

We learned the contents of this appendix from the referee of our previous paper \cite{IOYanok1}.

Let
\be
x_n:= \frac{A_{n+1}}{A_n}, \qq y_n:= \frac{B_{n+1}}{B_n}.
\ee
Then \eqref{ABrec} leads to
\bel{xyrec}
x_n+\frac{1}{x_{n-1}} = \frac{2\, n\, \underline{g}{}_s}{1-A_n B_n}, \qq
y_n + \frac{1}{y_{n-1}} = \frac{2 \, n\, \underline{g}{}_s}{1 - A_n B_n}.
\ee
By taking the difference of these two equations and shifting $n$ by one, we find
\be
x_{n+1} - y_{n+1} + \frac{1}{x_n} - \frac{1}{y_n}
= x_{n+1} - y_{n+1} + \frac{y_n - x_n}{x_n y_n} = 0.
\ee
Hence
\bel{xyrec2}
\frac{x_n y_n}{y_n-x_n} = \frac{1}{y_{n+1} - x_{n+1}}.
\ee
Also,  \eqref{ABrec2} leads to
\be
y_n - x_n = \frac{2\, M \, \underline{g}{}_s}{A_n B_n}.
\ee
Hence
\bel{ABtoxy}
A_n B_n = \frac{2\, M\, \underline{g}{}_s}{y_n - x_n}.
\ee
By multiplying $x_{n-1}$ to the first equation of \eqref{xyrec} and using \eqref{ABtoxy},
we find
\be
1 + x_n x_{n-1}
= \frac{2\, n\, \underline{g}{}_s x_{n-1}}
{\dis 1 - \frac{2\, M\, \underline{g}{}_s}{y_n - x_n}},
\ee
i.e., 
\be
1 - \frac{2\, M\, \underline{g}{}_s}{y_n - x_n} = \frac{2\, n \, \underline{g}{}_s x_{n-1}}{1+x_n x_{n-1}}.
\ee
From this equation, we have
\be
y_n =x_n + 2\, M\, \underline{g}{}_s 
\left( 1 - \frac{2\, n \, \underline{g}{}_s x_{n-1}}{1+x_n x_{n-1}} \right)^{-1}.
\ee
By substituting this into \eqref{xyrec2} and with some work, we obtain the following recursion relation
for $x_n$:
\be
\frac{2(n+1) \underline{g}{}_s}{1+x_n x_{n+1}} + \frac{2\, n \, \underline{g}{}_s}{1+x_n x_{n-1}}
= - x_n + \frac{1}{x_n} + 2\, n\, \underline{g}{}_s - 2\, M\, \underline{g}{}_s.
\ee
Similarly, the recursion relation for $y_n$ can be obtained:
\be
\frac{2(n+1) \underline{g}{}_s}{1+y_n y_{n+1}} + \frac{2\, n \, \underline{g}{}_s}{1+y_n y_{n-1}}
= - y_n + \frac{1}{y_n} + 2\, n\, \underline{g}{}_s + 2\, M\, \underline{g}{}_s.
\ee
These are the alternate discrete Painlev\'{e} II  equations  \cite{FGR93,NSKGR96}. By comparing
these equations with the alt-dPII equation  \cite[eq.(1.3)]{NSKGR96}
\be
\frac{\tilde{z}_n}{\tilde{x}_{n+1} \tilde{x}_n+1} + \frac{\tilde{z}_{n-1}}{\tilde{x}_n \tilde{x}_{n-1}+1}
= - \tilde{x}_n + \frac{1}{\tilde{x}_n} + \tilde{z}_n + \tilde{\mu},
\ee
where $\tilde{z}_n = \tilde{a} n + \tilde{b}$, we find
\be
\tilde{z}_n = \tilde{a} n+ \tilde{b}=2(n+1) \underline{g}{}_s, \qq
\tilde{a}=\tilde{b}=2\, \underline{g}{}_s,
\ee
$\tilde{\mu}=2(-M-1) \underline{g}{}_s$ for $\tilde{x}_n = x_n$ 
and $\tilde{\mu}=2(M-1) \underline{g}{}_s$
for $\tilde{x}_n = y_n$.
Note that the value of the parameter $\tilde{\mu}$ for $x_n$ and that for $y_n$ are different.

We also mention that our solutions 
\be
x_n = \frac{K^{(n+1)}_{M+1} K^{(n)}_M}{K^{(n+1)}_M K^{(n)}_{M+1}}, \qq
y_n = \frac{K^{(n+1)}_{M-1} K^{(n)}_M}{K^{(n+1)}_M K^{(n)}_{M-1}}
\ee
belong to a class of the Casorati determinant solutions to the alt-dPII considered in \cite{NSKGR96}.

\section{$A_{n,k}(M)$}
\label{AnkM}

The coefficients $A_{n,k}(M)$ are introduced in \eqref{Anexp}. 
For $0 \leq k \leq 4$, they are given in \eqref{An4}. For $5 \leq k \leq 11$, they are determined
as follows.
\be
\begin{split}
A_{n,5}(M) &= -n (2M+1)(2M-1) \cr
& \times \Bigl\{ n^4 (2M-3)(2M-5)(2M-7) \cr
& \qq -10 \, n^2 (2M+3)(2M-5)(2M-7)  \cr
& \qq + 9 (2M+5)(2M+3)(2M-3) \Bigr\},
\end{split}
\ee
\be
\begin{split}
A_{n,6}(M) &= n^2 (2M+1)(2M-1) \cr
& \times \Bigl\{ n^4 (2M-3)(2M-5)(2M-7)(2M-9) \cr
& \qq - 20\, n^2(2M+3)(2M-5)(2M-7)(2M-9) \cr
& \qq + 8 (2M+3)(2M-3)(32\, M^2 - 84 \, M - 395 ) \Bigr\},
\end{split}
\ee
\be
\begin{split}
A_{n,7}(M) &=
-n (2M+1)(2M-1) \cr
& \times \Bigl\{
n^6 (2M-3)(2M-5)(2M-7)(2M-9)(2M-11) \cr
& \qq - 35 \, n^4 ( 2M+3)(2M-5)(2M-7)(2M-9)(2M-11) \cr
& \qq + 7\, n^2 (2M+3)(2M-3) (2M-11)(148\, M^2 - 456 \, M - 1945) \cr
& \qq - 225 (2M+7)(2M+5)(2M+3)(2M-3)(2M-5) \Bigr\},
\end{split}
\ee
\be
\begin{split}
A_{n,8} &=
n^2(2M+1)(2M-1) \cr
& \times \Bigl\{
n^6 ( 2M-3)(2M-5) (2M-7)(2M-9)(2M-11)(2M-13) \cr
&- 56 n^4 (2M+3)(2M-5)(2M-7)(2M-9)(2M-11)(2M-13) \cr
& + 112\, n^2(2M+3)(2M-3)(2M-11)(2M-13)(28M^2-96M-385) \cr
& - 4608 (2M+5)(2M+3)(2M-3)(4M^3-29M^2-61M+266)
\Bigr\},
\end{split}
\ee
\be
\begin{split}
A_{n,9} &=-
n(2M+1)(2M-1) \cr
& \times \Bigl\{
n^8 (2M-3)(2M-5) (2M-7)(2M-9)(2M-11)(2M-13)(2M-15) \cr
& - 84 \, n^6 (2M+3)(2M-5)(2M-7)(2M-9)(2M-11)(2M-13)(2M-15) \cr
& +42 \, n^4 (2M+3)(2M-3)(2M-11)(2M-13)(2M-15)(188M^2-696M-2675) \cr
& - 4 \, n^2 (2M+3)(2M-3) \cr
& \qq \times (103328\, M^5
-1402128\,M^4+1055408\,M^3+32388552\,M^2 \cr
& \qq \qq -9301990\,M
-144452385) \cr
&+11025(2M+9)(2M+7)(2M+5)(2M+3)(2M-3)(2M-5)(2M-7)
\Bigr\},
\end{split}
\ee
\be
\begin{split}
&A_{n,10}(M) \cr
&=n^2(2\, M+1)(2\, M-1) \cr
&\times \Bigl\{
n^8  ( 2\,M-3 ) ( 2\,M-5)  ( 2\,M-7 )  
( 2\,M-9 )  ( 2\,M-11 )  \cr
& \qq \qq \times ( 2\,M-13 ) ( 2\,M-15 )  ( 2\,M-17 )  
\cr
&\qq - 120\, n^6 ( 2\,M+3 )  ( 2\,M-5 )( 2\,M-7 )( 2\,M-9 )    ( 2\,M-11 ) 
  \cr
 &\qq \qq \times     ( 2\,M-13 )( 2\,M-15 )   ( 2\,M-17 )
\cr
&\qq + 1344\,n^4  ( 2\,M+3 )( 2\,M-3 ) ( 2\,M-11 )( 2\,M-13 )  ( 2\,M-15 )     \cr
& \qq \qq \times  ( 2\,M-17 )  ( 13\,M^{2}-51\,M-190 ) 
\cr
&\qq -320\, n^2 ( 2\,M+3 )( 2\,M-3 )  ( 2\,M-17 ) \cr
& \qq \qq \times 
 ( 5248\,M^{5}-76608\,M^{4}+91048\,M^{3}+1797012\,M^{2} \cr
 & \qq \qq \qq -675770\,M-8008245 ) 
 \cr
&\qq +1152\, ( 2\,M+5 )  ( 2\,M+3 )  ( 2\,M-3 )  ( 2\,M-5 )  \cr
& \qq \qq \times ( 2048\,M^{4}-14960\,M^{3}-125420\,M^{2}+166700\,M+1176777 ) 
\Bigr\},
\end{split}
\ee
\be
\begin{split}
&A_{n,11}(M)\cr
&=
-n(2\, M+1)(2\, M-1) \cr
&\times \Bigl\{
n^{10}   ( 2\,M-3 )  ( 2\,M-5 )  ( 2\,M-7 )  ( 2\,M-9 ) ( 2\,M-11 )  \cr
& \qq \qq \times ( 2\,M-13 )  ( 2\,M-15 )  ( 2\,M-17 )  ( 2\,M-19 )
\cr
&\qq -165\, n^8 ( 2\,M+3 )  ( 2\,M-5 ) 
 ( 2\,M-7 )  ( 2\,M-9 )  ( 2\,M-11 )  \cr
 & \qq \qq \times ( 2\,M-13 )  ( 2\,M-15 )  ( 2\,M-17 )  ( 2\,M-19 ) 
\cr
&\qq + 462\,n^6   ( 2\,M+3 )  ( 2\,M-3 ) 
 ( 2\,M-11 )  ( 2\,M- 13 )  ( 2\,M-15 ) \cr
 & \qq \qq \times  ( 2\,M-17 )  ( 2\,M-19 )
( 76\,M^{2}-312\,M-1135 ) 
\cr
&\qq -110\, n^4   ( 2\,M+3 )  ( 2\,M-3 )  ( 2\,M-17 )  ( 2\,M-19 ) \cr
& \qq \qq \times ( 50272\,M^{5}-775152\,M^{4}
+1191952\,M^{3} \cr
& \qq \qq \qq +18444888\,M^{2} -8203850\,M-82219095 ) 
\cr
&\qq + 99\, n^2 ( 2\,M+5 )  ( 2\,M+3 )  ( 2\,M-3 )  \cr
& \qq \qq \times ( 683456\,M^{6}-14938112\,M^{5}+53795120\,M^{4}+
483527680\,M^{3} \cr
& \qq \qq \qq -1617082156\,M^{2}-3665527808\,M+10907967105 ) 
\cr
&\qq -893025\, ( 2\,M+11 )  ( 2\,M+9 )  ( 2\,M+7)  ( 2\,M+5 )  
( 2\,M+3 )  \cr
& \qq \qq \times( 2\,M-3 )  ( 2\,M-5 )  ( 2\,M-7 )  ( 2\,M-9 )
\Bigr\}.
\end{split}
\ee

\section{$\widetilde{J}_j(M)$}

\label{JjM}

We give a proof of \eqref{JjMcos}.
Recall that the infinite sum \eqref{JjMdef} is defined by
\be
\widetilde{J}_j(M) = (2j+1) \lim_{N \rightarrow \infty} \left( \sum_{k=0}^N 
\frac{(-1)^k \widetilde{\mathcal{C}}_{k+j}(M)}{k!\, (k+2j+1)!} \right).
\ee
From the definition \eqref{tCkM}, 
the coefficients $\widetilde{\mathcal{C}}_{k+j}(M)$ obey the following relations:
\be
\begin{split}
\bigl( M^2 -(j+1/2)^2 \bigr) \widetilde{\mathcal{C}}_{j+k}(M)
&= \Bigl[ M^2 - (k+j+1/2)^2 + k(k+2j+1) \Bigr]\widetilde{\mathcal{C}}_{j+k}(M) \cr
&= \widetilde{\mathcal{C}}_{j+k+1}(M) + k(k+2j+1) \widetilde{\mathcal{C}}_{j+k}(M).
\end{split}
\ee
Using these recursion relations, we can see that
\be
\bigl( M^2 -(j+1/2)^2 \bigr) \sum_{k=0}^N 
\frac{(-1)^k \widetilde{\mathcal{C}}_{k+j}(M)}{k!\, (k+2j+1)!} 
= \frac{(-1)^N \widetilde{\mathcal{C}}_{N+j+1}(M)}{N! \, (N+2j+1)!}.
\ee
Hence
\be
\sum_{k=0}^N 
\frac{(-1)^k \widetilde{\mathcal{C}}_{k+j}(M)}{k!\, (k+2j+1)!} 
=\frac{(-1)^{j+1} \mathcal{P}_{N,j}}{\{M^2-(j+1/2)^2 \}}
\prod_{k=1}^{N+j+1}\left[ 1 - \frac{M^2}{(k-1/2)^2} \right],
\ee
where
\be
\begin{split}
\mathcal{P}_{N,j}&:=\frac{1}{N! \, (N+2j+1)!}\prod_{k=1}^{N+j+1} (k-1/2)^2 \cr
&= \frac{1}{N!\, (N+2j+1)!} \left[ \frac{(2N+2j+2)!}{2^{2N+2j+2} \, (N+j+1)!} \right]^2.
\end{split}
\ee
We can see that
\be
\lim_{N \rightarrow \infty} \mathcal{P}_{N,j} = \frac{1}{\pi},
\ee
and
\be
\lim_{N \rightarrow \infty} \prod_{k=1}^{N+j+1}\left[ 1 - \frac{M^2}{(k-1/2)^2} \right]
= \cos (\pi M).
\ee
Therefore
\be
\lim_{N \rightarrow \infty} \left( \sum_{k=0}^N 
\frac{(-1)^k \widetilde{\mathcal{C}}_{k+j}(M)}{k!\, (k+2j+1)!} \right)
= \frac{(-1)^{j+1}}{\pi\, \{ M^2 - (j+1/2)^2 \}} \cos (\pi M),
\ee
which completes the proof of \eqref{JjMcos}.

Note that
\be
\widetilde{\mathcal{C}}_k(0) = (-1)^k \prod_{j=1}^k \left( \frac{2j-1}{2} \right)^2
= (-1)^k \frac{((2k)!)^2}{2^{4k}\, (k!)^2}.
\ee
Then we have
\be
\begin{split}
\widetilde{J}_j(0) &= (-1)^j (2j+1) \sum_{k=0}^{\infty}
\frac{[(2k+2j)!]^2}{2^{4(k+j)}\, k! (k+2j+1)! [(k+j)!]^2} \cr
&= \frac{ (-1)^{j} 4}{(2j+1) \pi}.
\end{split}
\ee

\section{Conformal block and its irregular limit}

\label{CBandIL}

The Coulomb-gas representation of the conformal block \cite{DF} can be interpreted as a matrix model \cite{MMM09093, IMO, EM09,MMS10, IYone}.
By considering ``colliding limits'' of the matrix model, 
the irregular conformal blocks can also be treated within the framework of the matrix model
\cite{EM10,IOYone,NR1,NR2,CR}.\footnote{
In \cite{MMS11}, the $N_f=0$ limit of the conformal block is considered and is interpreted as the
$\beta$-deformed GWW model.}
Following \cite{IOYone}, we review the matrix models
corresponding to the $SU(2)$ gauge theory with $N_f=2,3,4$ cases. Here we adopt 
integration contours slightly different from those found in \cite{IOYone}.
With these modifications, it is easy to see the relation between a special case of $N_f=2$ model and the unitary matrix model
with the logarithmic potential considered in Section \ref{UMMwLog}.

\subsection{Conformal block and $SU(2)$ gauge theory with $N_f=4$}

For simplicity, we assume that the parameter $q_0$ is real and $0 < q_0 < 1$.
For the matrix model \eqref{TCMM}, let us specify 
the integration contours to the intervals: $C_L^{(4)}=[0,q_0]$, $C_R^{(4)}=[1,\infty]$,  the potential
$W^{(4)}(z)$ to
\be
W^{(4)}(z) = \alpha_1 \log |z| + \alpha_2 \log|z-q_0|
+ \alpha_3 \log |z-1|,
\ee
and the normalization constant $\mathcal{C}^{(4)}$ to
\be
\mathcal{C}^{(4)} = q_0^{(1/2) \alpha_1 \alpha_2} (1-q_0)^{(1/2) \alpha_2 \alpha_3}.
\ee
Here the superscript $4$ in parenthesis represents the number of flavors of the corresponding gauge theory.
The matrix model \eqref{TCMM} in this case becomes
\begin{align}\label{3penner}
Z^{(4)} =& q_0^{(1/2) \alpha_1 \alpha_2} (1 - q_0)^{(1/2) \alpha_2 \alpha_3}
\left( \prod_{I=1}^N \int_{C^{(4)}_I} dw_I \right) \Delta(w)^{2\beta}\nonumber\\
&\times \prod_{I=1}^N \left| w_I \right|^{\sqrt{\beta} \alpha_1}\left| w_I - q_0 \right|^{\sqrt{\beta} \alpha_2} \left| w_I - 1 \right|^{\sqrt{\beta} \alpha_3}.
\end{align}
Here $C^{(4)}_I=C^{(4)}_L$ for $1 \leq I \leq N_L$ and $C^{(4)}_I=C^{(4)}_R$ for $N_L+1 \leq I \leq N$.

It is a Coulomb gas representation of the conformal block $\mathcal{F}$:
\be
Z^{(4)}(N_L, N_R)=\widetilde{C}_{43}{}^I C_{21}{}^I
\mathcal{F}(q_0 \, | \, c; \Delta_1, \Delta_2, \Delta_3, \Delta_4; \Delta_I),
\ee
\be
\mathcal{F}
= q_0^{\Delta_I - \Delta_1 - \Delta_2}
\left( 1 + \frac{(\Delta_I+\Delta_2 - \Delta_1)(\Delta_I  + \Delta_3- \Delta_4)}{2 \, \Delta_I}
q_0 + O(q_0^2) \right),
\ee
where the central charge is given by
$c = 1 - 6 Q_{E}^2$ with $Q_E =\sqrt{\beta} - 1/\sqrt{\beta}$, and the scaling dimensions are
\be
\Delta_i = \frac{1}{4}\alpha_i(\alpha_i-2Q_E), \qq
(i=1,2,3,4), \qq \Delta_I = \frac{1}{4} \alpha_I(\alpha_I-2Q_E).
\ee
Here  $\alpha_4=-2 Q_E - 2 \sqrt{\beta} N
- \alpha_1 - \alpha_2- \alpha_3$ and
$\alpha_I=\alpha_1 + \alpha_2 + 2 \sqrt{\beta} N_L$. The OPE coefficients $C_{21}{}^I$ and
$\widetilde{C}_{43}{}^I$ in the Coulomb gas model 
are respectively given by the Selberg integrals:
\be
\begin{split}
C_{21}{}^I
&=S_{N_L}(1+\sqrt{\beta} \alpha_1, 1 + \sqrt{\beta} \alpha_2, \beta) \cr
&= \prod_{J=1}^{N_L}
\frac{\Gamma(1 + j \beta) \Gamma( \sqrt{\beta}(\alpha_1 - Q_E) + j \beta)
\Gamma( \sqrt{\beta} (\alpha_2 - Q_E)+j \beta)}
{\Gamma(1+\beta) \Gamma(\sqrt{\beta}(\alpha_I - Q_E) - j \beta+1)},
\end{split}
\ee
\be
\begin{split}
\widetilde{C}_{43}{}^I
&=S_{N_R}(1+\sqrt{\beta} \alpha_3, 1 + \sqrt{\beta} \alpha_4, \beta) \cr
&= \prod_{J=1}^{N_R}
\frac{\Gamma(1 + j \beta) \Gamma( \sqrt{\beta}(\alpha_3 - Q_E) + j \beta)
\Gamma( \sqrt{\beta} (\alpha_4 - Q_E)+j \beta)}
{\Gamma(1+\beta) \Gamma(\sqrt{\beta}(Q_E - \alpha_I) - j \beta+1)}.
\end{split}
\ee

This matrix model contains seven parameters 
\begin{align}\label{Nf4param}
(\sqrt{\beta},N_L, \alpha_1 , \alpha_2 ; N_R, \alpha_3, \alpha_4 ),\quad N \equiv N_L + N_R
\end{align}
with one constraint
\begin{align}\label{constrain}
\alpha_1 + \alpha_2 + \alpha_3 + \alpha_4 + 2 \sqrt{\beta} N = 2 Q_E.
\end{align}
These are related to six unconstrained parameters of $4$d  $SU(2)$ gauge theory with $N_f=4$
\begin{align}
\frac{\epsilon_1}{g_s},~ \frac{\mathfrak{a}}{g_s},~ \frac{m_1}{g_s},~ \frac{m_2}{g_s},~ \frac{m_3}{g_s},~ \frac{m_4}{g_s},
\end{align}
by the $0$d-$4$d dictionary \cite{IO5}\footnote{Here we have renamed the mass parameters such that
	the ordering of mass to infinity limit is natural.}:
\begin{align}
\sqrt{\beta} N_L &= \frac{\mathfrak{a} - m_2}{g_s} ,& 
\sqrt{\beta} N_R &= -\frac{\mathfrak{a} + m_1}{g_s}, \nonumber\\
\alpha_1 &= \frac{1}{g_s} \left( m_2 - m_4 + \epsilon \right) , & 
\alpha_2 &= \frac{1}{g_s} \left( m_2 + m_4 \right), \label{dict} \\
\alpha_3 &= \frac{1}{g_s} \left( m_1 + m_3 \right), & 
\alpha_4 &= \frac{1}{g_s}\left( m_1 - m_3 + \epsilon \right) . \nonumber
\end{align}
The omega background parameters $\epsilon_{1,2}$ are related to $\beta$ as $\epsilon_1 = \sqrt{\beta} \, g_s$
and $\epsilon_2 = - g_s/\sqrt{\beta}$. Hence $g_s^2 = - \epsilon_1 \, \epsilon_2$ and $\epsilon \equiv
\epsilon_1 + \epsilon_2 = Q_E \, g_s$.
The cross ratio $q_0$ is identified with the exponentiated ultraviolet gauge coupling constant
$q_0 \equiv e^{\im \pi \tau_0}$, 
$\tau_0 \equiv (\theta_0/\pi) + 8\,  \pi \, \im/g_0^2$.

\subsection{Irregular limit to $N_f=3$}

In order to obtain well-defined $\beta$-deformed matrix models corresponding to the $N_f = 3,\,2$ gauge theories, we should take the analytic continuation of \eqref{3penner}.

\begin{figure}
	\centering
	\includegraphics[width = 10cm,bb=0 0 202 117]{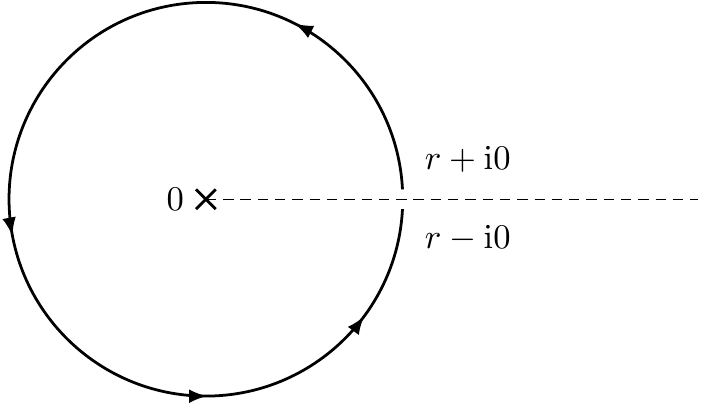}
	\caption{The integration contour $C(r)$ which starts at $r+\im 0$ and ends at $r-\im 0$. The cross 
	and the dashed line denote the branch point and the cut respectively.} 
	\label{fig1a} 
\end{figure}

Let us consider an integral of the form
\be
\int \de w\, w^{\alpha} \left( \sum_{n=0}^{\infty} c_n w^n \right),
\ee
for the integration contour $C^{(4)}_L$ or $C^{(4)}_R$.
Here $c_n$ are constants and  $\alpha$ a complex parameter which is not an integer.  
Let $C(r)$ be the contour which is an arc of radius $r$ in the complex $w$-plane,
running counterclockwise from $r+\im 0$ to $r- \im 0$. It avoids the cut of $w^{\alpha}$ along positive 
$\mathrm{Re}\, w$-axis (See Fig.\ref{fig1a}).

For $\mathrm{Re} \, \alpha>0$, the integral over 
the interval $C^{(4)}_L=[0,q_0]$ can be converted to a complex integral along $C(r)$ with $r=q_0$ by
\be
\int_0^{q_0} \de w\, w^{\alpha} \left( \sum_{n=0}^{\infty} c_n w^n \right)
= \frac{1}{(\ex^{2\pi \im \alpha} - 1)} \int_{C(q_0)} 
\de w\, w^{\alpha} \left( \sum_{n=0}^{\infty} c_n w^n \right).
\ee
Similarly, for $\mathrm{Re}\, \alpha<0$, the integral over $C^{(4)}_R=[1,\infty]$ becomes
the following complex integral along $C(r)$ with $r=1$:
\be
\int_1^{\infty} \de w\, w^{\alpha} \left( \sum_{n=0}^{\infty} c_n w^{-n} \right)
= \frac{1}{(1- \ex^{2\pi \im \alpha})}
\int_{C(1)} \de w\, w^{\alpha} \left( \sum_{n=0}^{\infty} c_n w^{-n} \right).
\ee

\begin{figure}
	\centering
	\includegraphics[width=10cm,bb=0 0 208 133]{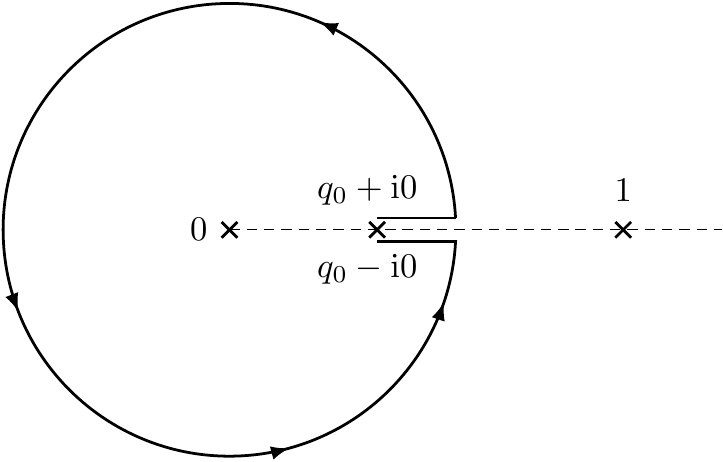}
	\caption{The integration contour $C_L(r_0)$ which is a deformation of $C(q_0)$. It starts at $q_0+\im 0$
	and ends at $q_0 - \im 0$. The radius of the arc is $r_0$.}
	\label{fig2a}
\end{figure}

Using these relations, we assume that \eqref{3penner} can be rewritten as
\begin{align}
Z^{(4)} =& q_0^{(1/2) \alpha_1 \alpha_2} (1 - q_0)^{(1/2) \alpha_2 \alpha_3} \, \mathcal{C}_L \, \mathcal{C}_R
\left( \prod_{I=1}^N \int_{C^{\prime (4)}_I} dw_I \right) \Delta(w)^{2\beta}\nonumber\\
&\times \prod_{I=1}^N w_I^{\sqrt{\beta} \alpha_{1+2}} \left( 1 - \frac{q_0}{w_I} \right)^{\sqrt{\beta} \alpha_2} \left( w_I - 1 \right)^{\sqrt{\beta} \alpha_3},
\end{align}
where $\mathcal{C}_L$ and $\mathcal{C}_R$ are the normalization factors. Up to phase factors, they are given by
\be
\mathcal{C}_L = \prod_{I=1}^{N_L} 
\Bigl( \ex^{2\pi \im (\sqrt{\beta} \alpha_1 + 2 (I-1) \beta)} - 1 \Bigr)^{-1},
\qq
\mathcal{C}_R = \prod_{J=1}^{N_R}
\Bigl( 1 - \ex^{- 2\pi \im ( \sqrt{\beta} \alpha_4 + 2(J-1) \beta)}\Bigr)^{-1}.
\ee
The contour $C^{\prime (4)}_I$ is the complex contour $C_{L}(r_0)$ (see Fig.\ref{fig2a}) for $1 \leq I \leq N_L$ and $C(1)$ for $N_L + 1 \leq I \leq N$. 
Here $r_0$ is an arbitrary real number satisfying $0 < q_0 < r_0<1$. 
The contour $C_L(r_0)$ is a deformation of $C(q_0)$.

The $N_f=3$ limit of the gauge theory is taken by $m_4 \rightarrow \infty$ with 
$\Lambda_3 \equiv 4\, q_0 \, m_4$ fixed. 
Using above dictionary \eqref{dict}, this corresponds to
the $q_0 \rightarrow 0$ limit
with $2 \, q_{03} \equiv q_0 \left( -\alpha_1 + \alpha_2 \right)$ and
$\alpha_{1+2}\equiv \alpha_1+\alpha_2$ kept finite. 
The parameter $q_{03}$
is related to the dynamical mass scale $\Lambda_3$ of the $N_f=3$ theory as
$q_{03}
= \Lambda_3/(4\, g_s)$.

Note that the out of the seven constraint parameters in \eqref{Nf4param}, six ones are kept intact:
\begin{align}
(\sqrt{\beta}, \alpha_{1+2} , N_L ; \alpha_3, \alpha_4, N_R).
\end{align}
The constraint \eqref{constrain} reduces to
\begin{align}
\alpha_{1+2} + \alpha_3 + \alpha_4 + 2 \sqrt{\beta} N  = 2 Q_E . \label{constNf3}
\end{align}
In \eqref{dict}, the dictionary for $\alpha_1$ and that for $\alpha_2$ get replaced with
\be
\alpha_{1+2} = \frac{2 m_2 + \epsilon}{g_s}.
\ee

\begin{figure}
	\centering
	\includegraphics[width=10cm,bb=0 0 208 132]{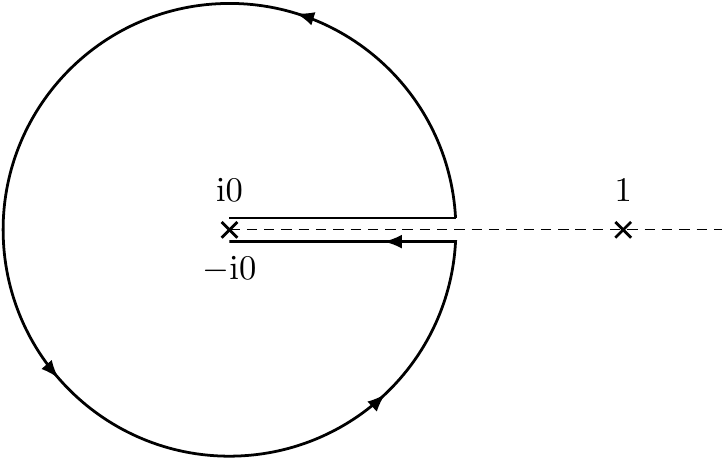}
	\caption{The integration contour $C_0(r_0)$. It starts at $\im 0$
	and ends at $- \im 0$. The radius of the arc is $r_0$.}
	\label{fig3a}
\end{figure}

We can then define the well-defined limit of the matrix model 
which corresponds to the $N_f=3$ limit of the gauge theory,
\begin{align}
Z^{(3)} &:=  \lim_{q_0 \rightarrow 0} 
\frac{1}{q_0^{(1/2)\alpha_1 \alpha_2} (1 - q_0)^{(1/2) \alpha_2 \alpha_3} \,
\mathcal{C}_L \, \mathcal{C}_R} Z^{(4)} \nonumber\\
&= \left( \prod_{I=1}^N \int_{C^{(3)}_I }d w_I \right) \Delta(w)^{2 \beta}   \prod_{I=1}^N w_I^{\sqrt{\beta} \alpha_{1+2}} \exp\left( - \frac{\sqrt{\beta} q_{03}}{w_I} \right) \left( w_I - 1 \right)^{\sqrt{\beta} \alpha_3}.
\end{align}
In this limit, the contour $C_L(r_0)$ goes to $C_0(r_0)$ which are shown in Fig.\ref{fig3a}.
The resultant contour $C_I^{(3)}$ is $C_0(r_0)$ for $1 \leq I \leq N_L$ and $C(1)$ for $N_L + 1 \leq I \leq N$.
The corresponding potential is given by
\be
W^{(3)}(z)= \alpha_{1+2} \log z - \frac{q_{03}}{z} + \alpha_3 \log(z-1).
\ee

\subsection{Irregular limit to $N_f=2$}

We can take the $N_f=2$ limit in the gauge theory subsequently after the $N_f = 3$ limit by $m_3 \rightarrow \infty$ with the dynamical scale $\Lambda_2 \equiv (m_3\, \Lambda_3)^{1/2}$ fixed. In the matrix model, this corresponds to the $q_{03} \rightarrow 0$ limit
with $q_{02}{}^2 \equiv (1/2)q_{03} \left( \alpha_3 - \alpha_4 \right)$
and $\alpha_{3+4} \equiv \alpha_3 + \alpha_4$ kept finite. 
In \eqref{Nf4param}, five parameters are kept intact:
\begin{align}
(\sqrt{\beta}, \alpha_{1+2}, N_L ; \alpha_{3+4}, N_R).
\end{align}
The momentum conservation \eqref{constNf3}
becomes
\begin{align}
\alpha_{1+2} + \alpha_{3+4} + 2 \sqrt{\beta} N  = 2\,  Q_E. \label{constNf2}
\end{align}
Now, the dictionary for $\alpha_3$ and that for $\alpha_4$ get replaced with
\begin{align}
\alpha_{3+4} = \frac{2 m_1 + \epsilon}{g_s}
\end{align}
in \eqref{dict} and $q_{02}$ is related to $\Lambda_2$ as $q_{02}=\Lambda_2/(2\, g_s)$.

\begin{figure}
	\centering
	\includegraphics[width = 10cm,bb=0 0 208 132]{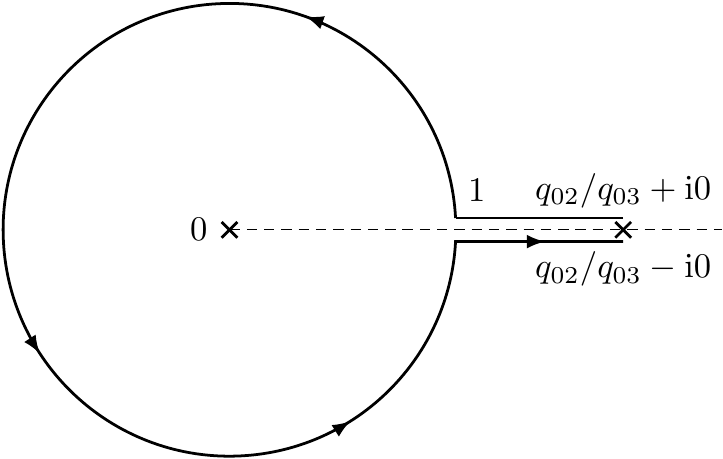}
	\caption{The integration contour $C_R(q_{02}/q_{03})$ which is a deformation of $C(q_{02}/q_{03})$.
	It starts at $q_{02}/q_{03}+\im 0$ and ends at $q_{02}/q_{03}-\im 0$.}
	\label{fig4a}
\end{figure}

\begin{figure}
	\centering
	\includegraphics[width = 10cm,bb=0 0 208 132]{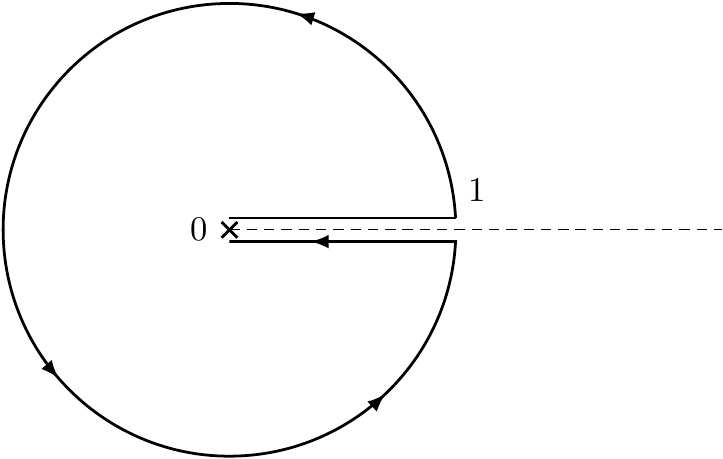}
	\caption{The integration contour $C_0$.}
	\label{fig5a}
\end{figure}

We rescale the integration variables as $w_I \rightarrow  (q_{03} / q_{02}) w_I$,
then we get
\begin{align}
Z^{(3)} = \mathcal{N}^{(3)} \left( \prod_{I=1}^N \int_{C^{\prime (3)}_I } d w_I \right)
\Delta(z)^{2 \beta} \prod_{I=1}^N w_I^{\sqrt{\beta} \alpha_{1+2}} \exp \left( - \frac{\sqrt{\beta} q_{02}}{w_I} \right) \left( 1 - \frac{q_{03}}{q_{02}} w_I \right)^{\sqrt{\beta} \alpha_{3}},
\end{align}
where the integration contour $C^{\prime (3)}_I$ is $C_0(r_0 \, q_{02}/q_{03})$ for $1 \leq I \leq N_L$
and $C_R(q_{02}/q_{03})$ (see Fig. \ref{fig4a}) for $N_L+1 \leq I \leq N$.
For simplicity, we set $r_0= q_{03}/q_{02}$. Then 
$C_0(r_0 \, q_{02}/q_{03}) = C_0(1) \equiv C_0$. (See Fig. \ref{fig5a}).
The normalization factor $\mathcal{N}^{(3)}$ is given by
\be
\mathcal{N}^{(3)}
= \ex^{\im \sqrt{\beta} N \alpha_3 \pi} 
\left( \frac{q_{03}}{q_{02}} \right)^{\sqrt{\beta} N M},
\ee 
where \be
\begin{split}
M&:=\alpha_{1+2} + \sqrt{\beta} N - Q_E  = - \alpha_{3+4} - \sqrt{\beta} N + Q_E \cr
&= \frac{1}{2} ( \alpha_{1+2} - \alpha_{3+4}) =\frac{(m_2-m_1)}{g_s}.
\end{split}
\ee

\begin{figure}
	\centering
	\includegraphics[width = 10cm,bb=0 0 208 132]{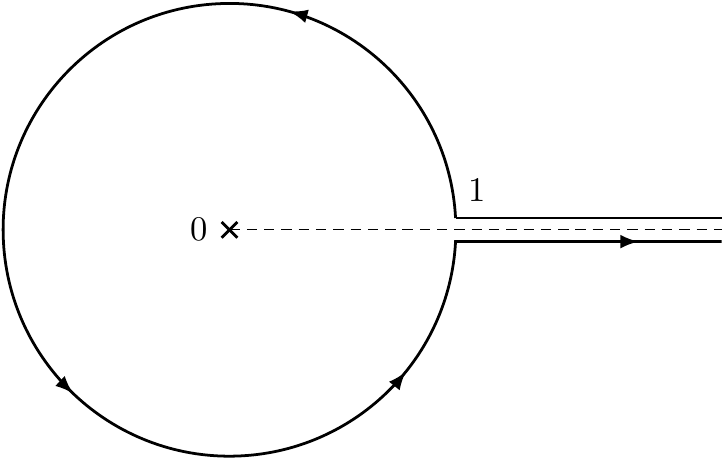}
	\caption{The integration contour $C_{\infty}$. This is a limit of $C_R(q_{02}/q_{03})$ (see Fig.\ref{fig5a}), obtained by sending $q_{02}/q_{03} \rightarrow + \infty$.}
	\label{fig6a}
\end{figure}

We define the $N_f = 2$ limit for the matrix model by
\begin{align} \label{Z2}
Z^{(2)} &\equiv \lim_{q_{03} \rightarrow 0} \frac{1}{\mathcal{N}^{(3)}} Z^{(3)} \nonumber\\
&=\left( \prod_{I=1}^N \int_{C^{(2)}_I } dw_I \right) \Delta(w)^{2 \beta} \prod_{I=1}^N w_I^{\sqrt{\beta} \alpha_{1+2}} 
\exp\left( - \sqrt{\beta} q_{02} \left( w_I + \frac{1}{w_I} \right) \right).
\end{align}
The resultant integration contours $C_I^{(2)}$ are $C_0$ for $1 \leq I \leq N_L$ and $C_\infty$ (see Fig.\ref{fig6a}) for $N_L + 1 \leq I \leq N$. The corresponding potential is given by
\bel{potNf2}
W^{(2)}(z) = - q_{02} \left( z + \frac{1}{z} \right) + \alpha_{1+2} \log z.
\ee

Note that if $\beta$ is a positive integer and $\sqrt{\beta} \alpha_{1+2} \in \mathbb{Z}$, 
then the cut for the integrand of \eqref{Z2} vanishes.
In this case, the open paths $C_0$ and $C_{\infty}$ can be deformed to the closed circles with unit radius
around the origin.
Thus, we can represent \eqref{Z2} as the multiple integral over the same unit circles.
In particular, for $\beta=1$ with $\alpha_{1+2} \in \mathbb{Z}$, 
we can identify $N_f = 2$ matrix model \eqref{Z2} with an unitary matrix model with a logarithmic potential.

To see the connection with the unitary matrix model explicitly, 
let us rewrite the partition function \eqref{Z2} as follows:
\bel{Z2U}
	Z^{(2)} = \left( \prod_{I=1}^N \int_{C_I^{(2)}} \frac{dw_I}{w_I} \right) \Delta(w)^{ \beta} \Delta(w^{-1})^{ \beta} \exp \left[ \sqrt{\beta} \sum_{I=1}^N
	\left\{ - q_{02} \left( w_I + \frac{1}{w_I} \right) + M \log w_I \right\}\right].
\ee
For $\beta=1$, it holds that $M = \alpha_{1+2} +N$ and $\alpha_{1+2} \in \mathbb{Z}$ is equivalent
to $M \in \mathbb{Z}$. Thus for this case with $\beta=1$, \eqref{Z2U} is indeed a unitary matrix model.
Note that in this case, the partition function $Z^{(2)}=Z^{(2)}(N_L, N_R)$
depends only on the sum $N=N_L+N_R$:
\be
Z^{(2)}(N_L, N_R) = Z^{(2)}(N,0) =: Z_{U(N)}.
\ee
So, their generating function \eqref{GTCMM} becomes
\be
\underline{Z}{}_{U(N)}=\underline{Z}^{(2)}(N; \mu_L, \mu_R) =\frac{(\mu_L+\mu_R)^N}{N!} Z_{U(N)}.
\ee
Hence, in gauge theory parameters, this does not depend on the vev $\mathfrak{a}$ of the Higgs field.
In the matrix model/gauge theory correspondence, the $N_f=2$ instanton partition function
is identified with $Z^{(2)}/(\mathcal{N}^{(2)} Z^{(2)}_L Z^{(2)}_R )$ (see \eqref{NekNf2}).
The dependence on $\mathfrak{a}$ comes from $Z^{(2)}_L(N_L) Z^{(2)}_R(N_R)$.

\section{Loop equations}
\label{LE}

In this section, we briefly review the finite $N$ loop equations for the $\beta$-deformed matrix models 
and their planar limit. See \cite{Dai, MM90, IM} for earlier literature at $\beta = 1$.
See also \cite{IMO, IY} for examples at higher rank extensions.

\subsection{Finite $N$ loop equations}

For the matrix model \eqref{TCMM} or \eqref{GTCMM}, we assume that the potential $W(z)$ has
the following property:
\be
\int \de^N w \frac{\partial}{\partial w_I} \left[ F(w) \Delta^{2\beta}(w) 
\exp\left( \sqrt{\beta} \sum_J W(w_J) \right) \right] = 0, \qq (I=1,2,\dotsm, N),
\ee
where $F(w)$ is a holomorphic function of $\{ w_J\}$  in a domain which includes the integration contours.

Let $\omega_N(w)$ be a finite $N$ resolvent defined by
\be
\omega_N (z) = \sqrt{\beta} \sum_{I=1}^N \frac{1}{z - w_I}.
\ee

It obeys the finite $N$ loop equation:
\bel{FLEq}
 \dket \Bigl( \omega_N(z) \Bigr)^2 \dbr + \left( W^\prime(z) + Q_E \frac{d}{d z} \right) 
 \Bigl\langle \!\! \Bigl\langle \omega_N(z) \Bigl \rangle \!\! \Bigl\rangle - f_N (z) = 0,
\ee
where
\be
f_N(z) = \dket \sqrt{\beta} \sum_{I=1}^N\frac{W^\prime(z) - W^\prime(w_I)}{z - w_I} \dbr.
\ee
Here $\langle\!\langle \dotsm \rangle\! \rangle$ means the average with respect to $Z(N_L,N_R)$
\eqref{TCMM} or $\underline{Z}(N; \mu_L, \mu_R)$ \eqref{GTCMM}.

We denote the power sum by
\be
p_{N,\ell}(w) = \sum_{I=1}^N w_I^{\ell}, \qq (\ell=1,2,\dotsm).
\ee
Then
\begin{align}
	\omega_N(z) =\sqrt{\beta} \sum_{\ell=0}^\infty \sum_{I=1}^N \frac{w_I{}^\ell}{z^{\ell + 1}} = \frac{\sqrt{\beta} N}{z} + \sqrt{\beta} \sum_{\ell =1}^\infty \frac{p_{N,\ell} (w)}{z^{\ell+1}}.
\end{align}

\subsection{Planar loop equation and Seiberg-Witten curve}
 
\label{AppPLEq}

In this subsection, we rescale the potential $W(z)$ of the matrix model by
$W(z) \rightarrow (1/g_s) W(z)$, so
\bel{PZ}
Z = \int \de^N w \, \Delta^{2\beta}(w)
\exp\left( \frac{\sqrt{\beta}}{g_s} \sum_{I=1}^N W(w_I) \right).
\ee
Let us consider the planar limit of the finite $N$ loop equation.
It is the limit of $N \rightarrow \infty$ and $g_s \rightarrow 0$ with
keeping the 't Hooft coupling $\widetilde{S}:=\sqrt{\beta} \, g_s N$ finite.

The (planar) resolvent $\omega(z)$ is defined by
\bel{presol}
\omega(z):= \lim_{g_s \rightarrow 0} g_s \, \langle \! \langle \omega_N(w) \rangle \! \rangle
= \lim_{g_s \rightarrow 0}
g_s \dket \sqrt{\beta} \sum_{I=1}^N\frac{1}{z - w_I} \dbr.
\ee
Here $\langle \! \langle \dotsm \rangle \! \rangle$ is the average with respect to \eqref{PZ}.
Also, let
\be
f(z):= \lim_{g_s \rightarrow 0} g_s
\dket \sqrt{\beta} \sum_{I=1}^N\frac{W^\prime(z) - W^\prime(w_I)}{z - w_I} \dbr.
\ee
In the planar limit, the finite $N$ loop equation \eqref{FLEq} turns into the planar loop equation
\be
\omega(z)^2 + W'(z) \omega(z) - f(z) = 0.
\ee
By introducing $y(z):=w(z) +(1/2)W'(z)$, this can be rewritten as
\bel{PlanarCurve}
y^2 = f(z) + \frac{1}{4} W'(z)^2.
\ee
In the matrix model/gauge theory correspondence, this spectral curve is identified with the Seiberg-Witten curve
of the corresponding gauge theory.

\section{Instanton expansion for $N_f=2$ matrix model}
\label{IEforN2MM}

Let us consider the instanton expansion (small $q_{02}$ expansion)
of the partition function for the $N_f=2$ model \eqref{Z2}. See \cite{IO5} and \cite{Zam, Ake, MMM09092,MMM09093, IMO,EM09, IYone, GT, IY}.
To consider the $q_{02}$ expansion, it is convenient to divide the $N$ integration variables $w_I$ ($I=1,2,\dotsm, N$) into two sets $\{ w_I \}_{I=1,2\dotsm, N_L}$ and $\{ w_{N_L+J} \}_{J=1,2, \dotsm, N_R}$
and rename the latter variables by $u_J:= w_{N_L+J}$ ($J=1,2,\dotsm, N_R$). Here $N=N_L + N_R$.

Furthermore, we rescale the integration variables as $w_I \rightarrow q_{02} \, w_I$ ($I=1,2,\dots, N_L$) 
and $u_J \rightarrow  (q_{02})^{-1} u_J$ ($J=1,2,\dotsm, N_R$).
In this rescaling, the integration contours $C_0$ and $C_{\infty}$ are also rescaled, but can be smoothly
deformed back to $C_0$ and $C_{\infty}$ respectively. Then, \eqref{Z2} becomes
\be
\begin{split}
Z^{(2)} &= \mathcal{N}^{(2)} \left(  \prod_{I=1}^{N_L} \int_{C_0} \de w_I \right) \Delta^{2\beta}(w)
\exp\left( \sqrt{\beta} \sum_{I=1}^{N_L} W_L(w_I) \right) \cr
& \times \left( \prod_{J=1}^{N_R}\int_{C_{\infty}} \de u_J \right) \Delta^{2\beta}(u) 
\exp\left( \sqrt{\beta} \sum_{J=1}^{N_R} W_R(u_J) \right) \cr
& \times \prod_{I=1}^{N_L} \prod_{J=1}^{N_R}
\left( 1 - q_{02}{}^2 \frac{w_I}{u_J} \right)^{2\beta}
\prod_{I=1}^{N_L} \exp\left( - \sqrt{\beta} q_{02}{}^2 w_I \right)
\prod_{J=1}^{N_R} \exp\left( - \sqrt{\beta} q_{02}{}^2 \frac{1}{u_J} \right),
\end{split}
\ee
where $\mathcal{N}^{(2)} = (q_{02})^{\sqrt{\beta} (N_L-N_R)M - 2 \beta N_L N_R}$,
\be
W_L(w) := - \frac{1}{w} + a_L \log w, \qq
W_R(u) := - u + a_R \log u,
\ee
with
\be
a_L :=\alpha_{1+2}=\frac{2\, m_2+\epsilon}{g_s}, \qq
a_R:=\alpha_{1+2} + 2 \sqrt{\beta} N_L = \frac{2 \, \mathfrak{a}+\epsilon}{g_s}.
\ee
Let us define the left and right partition functions by
\bel{partL}
	Z_L^{(2)}=Z_{L}^{(2)}(N_L) =  \int_{C_0} d^{N_L} w   \, \Delta^{2\beta}(w) 
	\exp \left( \sqrt{\beta} \sum_{I=1}^{N_L} W_L(w_I)\right), 
\ee
\bel{partR}
	Z_{R}^{(2)} =  Z_R^{(2)}(N_R)=\int_{C_{\infty}} d^{N_R} u   \, \Delta^{2\beta}(u) 
	\exp \left( \sqrt{\beta} \sum_{J=1}^{N_R} W_{R} (u_J)\right).
\ee
Then
\be
\begin{split}
Z^{(2)} &= \mathcal{N}^{(2)} \, Z_L^{(2)} \, Z_R^{(2)} \cr
& \times \dket \prod_{I=1}^{N_L} \prod_{J=1}^{N_R} \left( 1 - q_{02}{}^2 \frac{w_I}{u_J} \right)^{2 \beta} \prod_{I=1}^{N_L} \exp \left( - \sqrt{\beta} q_{02}{}^2 w_I \right) \prod_{J=1}^{N_R} \exp \left( - \sqrt{\beta} q_{02}{}^2 \frac{1}{u_J} \right) \dbr_{N_L,N_R},
\end{split}
\ee
Here $\langle\! \langle f(w,u) \rangle \!\rangle_{N_L,N_R}$ means the expectation value of $f(w,u)$ 
with respect to $Z_L^{(2)} Z_R^{(2)}$.

Now, $Z^{(2)}/( \mathcal{N}^{(2)} Z^{(2)}_L Z^{(2)}_R)$ has the following $q_{02}$ expansion
\bel{NekNf2}
\frac{Z^{(2)}}{\mathcal{N}^{(2)} Z^{(2)}_L Z^{(2)}_R}
= 1 + \sum_{\ell=1}^{\infty} q_{02}{}^{2\ell} \mathcal{A}_{\ell},
\ee
which is identified with the instanton part of the $N_f=2$ Nekrasov function.

For simplicity, we evaluate the first coefficient $\mathcal{A}_1$ 
\bel{instA1}
\mathcal{A}_{1} = -2 \beta \dket \sum_{I=1}^{N_L} w_I \dbr_{N_L}  \dket \sum_{J=1}^{N_R} \frac{1}{u_J} \dbr_{N_R} - \sqrt{\beta} \dket \sum_{I=1}^{N_L} w_I \dbr_{N_L} - \sqrt{\beta} \dket \sum_{J=1}^{N_R} \frac{1}{u_J} \dbr_{N_R}
\ee
with help of the finite $N$ loop equations.
Let us introduce the finite $N$ resolvent for $\{ w_I \}$ and that for $\{ u_J \}$ respectively by
\be
w_{N_L}(z):=\sqrt{\beta} \sum_{I=1}^{N_L} \frac{1}{z-w_I}, \qq
w_{N_R}(z):=\sqrt{\beta} \sum_{J=1}^{N_R} \frac{1}{z-u_J}.
\ee
They respectively satisfy the finite $N$ loop equations
\be
 \dket \Bigl( \omega_{N_{\mathcal{I}}} (z) \Bigr)^2 \dbr_{N_{\mathcal{I}}} + 
 \left( W_{\mathcal{I}}^\prime(z) + Q_E \frac{d}{d z} \right) 
 \Bigl\langle \!\! \Bigl\langle \omega_{N_{\mathcal{I}}}(z) \Bigl \rangle \!\! \Bigl\rangle_{N_{\mathcal{I}}}
 - f_{N_{\mathcal{I}}} (z) = 0, \qq(\mathcal{I}=L, R),
\ee
where $\langle\!\langle \dotsm \rangle\!\rangle_{N_{\mathcal{I}}}$ is the average with respect to $Z_{N_{\mathcal{I}}}$, and
\be
f_{N_L}(z) := \dket \sqrt{\beta} \sum_{I=1}^{N_L} \frac{W_L^\prime(z) - W_L^\prime(w_I)}{z - w_I} \dbr_{N_L},
\ee
\be
f_{N_R}(z) := \dket \sqrt{\beta} \sum_{J=1}^{N_R} \frac{W_R^\prime(z) - W_R^\prime(u_J)}{z - u_J} \dbr_{N_R}.
\ee
They are evaluated as
\be
	f_{N_L}(z) = - \frac{1}{z^2} \sqrt{\beta} \dket \sum_{I=1}^{N_L} \frac{1}{w_I} \dbr_{N_L} 
	- \frac{1}{z} \left( \sqrt{\beta} \, a_L  \dket \sum_{I=1}^{N_L} \frac{1}{w_I}\dbr_{N_L} + \sqrt{\beta} \dket \sum_{I=1}^{N_L} \frac{1}{w_I{}^2} \dbr_{N_L} \right),
\ee
\be
 f_{N_R}(z) = -\frac{1}{z}\sqrt{\beta} \, a_R \dket\sum_{J=1}^{N_R} \frac{1}{u_J} \dbr_{N_R} .
\ee
By comparing the $\mathcal{O}(z^{-2}) , \mathcal{O}(z^{-3})$ terms of the loop equation
for the left resolvent $\omega_{N_L}(z)$, we find
\begin{align}
	\sqrt{\beta} \dket \sum_{I=1}^{N_L} \frac{1}{w_I} \dbr_{N_L} 
	&= -\sqrt{\beta} N_L (\sqrt{\beta} N_L + a_L - Q_E) 
	= -\frac{(\mathfrak{a}+m_2)(\mathfrak{a}-m_2)}{g_s^2},\\
	\sqrt{\beta}\Bigl\langle \!\! \Bigl\langle p_{N_L,1} (w) \Bigr\rangle \!\! \Bigr\rangle_{N_L} 
	&=- \frac{\sqrt{\beta} N_L}{ 2 \sqrt{\beta} N_L + a_L - 2 Q_E} 
	=- \frac{(\mathfrak{a} - m_2)}{(2 \mathfrak{a} - \epsilon)}. \label{pL1}
\end{align}
In addition, from $O(z^{-1})$ terms, we have
\begin{align}
	\sqrt{\beta} \dket \sum_{I=1}^{N_L} \frac{1}{w_I{}^2} \dbr_{N_L} 
	&= - a_L \sqrt{\beta} \dket \sum_{I=1}^{N_L} \frac{1}{w_I} \dbr_{N_L} 
	= \sqrt{\beta} N_L  (\sqrt{\beta}N_L - a_L - Q_E) a_L\nonumber\\
	&= \frac{(\mathfrak{a} - m_2)(\mathfrak{a}+m_2)(2 \, m_2 + \epsilon)}{g_s^3}.
	\end{align}
Similarly, from the loop equation for $\omega_{N_R}(w)$, we obtain
\begin{align}
	\sqrt{\beta} \dket\sum_{I=1}^{N_R} \frac{1}{u_J} \dbr_{N_R} &= \frac{\sqrt{\beta} N_R}{a_R} 
	= -\frac{(\mathfrak{a} + m_1)}{(2\, \mathfrak{a} + \epsilon)}, \label{pRm1} \\
	\sqrt{\beta} \Bigl\langle \!\! \Bigl\langle p_{N_R,1} (u) 
	\Bigr\rangle \!\! \Bigr\rangle_{N_R} &= \sqrt{\beta} N_R \, (\sqrt{\beta} N_R + a_R - Q_E) 
	= - \frac{(\mathfrak{a}-m_1)(\mathfrak{a}+m_1)}{g_s^2},\\
	\sqrt{\beta} \Bigl\langle \!\! \Bigl\langle p_{N_R,2} (u) \Bigr\rangle \!\! \Bigr\rangle_{N_R} 
	&= (2 \sqrt{\beta} N_R + a_R - 2 \, Q_E) 
	\sqrt{\beta} \Bigl\langle \!\! \Bigl\langle p_{N_R,1} (u)  \Bigr\rangle \!\! \Bigr\rangle_{N_R}\nonumber\\
	&= \sqrt{\beta} N_R \, (\sqrt{\beta} N_R + a_R - Q_E) (2 \sqrt{\beta} N_R + a_R -2 \, Q_E)\nonumber\\
	&=\frac{(\mathfrak{a}-m_1)(\mathfrak{a}+m_1) (2 \, m_1 + \epsilon)}{g_s^3}.
\end{align}

By substituting \eqref{pL1} and \eqref{pRm1} into \eqref{instA1}, 
we find
\be
\mathcal{A}_{1}
= \frac{(\mathfrak{a} + m_1)(\mathfrak{a} + m_2)}
{2\, \mathfrak{a} (2\, \mathfrak{a} + \epsilon)} 
+ \frac{(\mathfrak{a} - m_1)(\mathfrak{a} - m_2)}
{2\, \mathfrak{a} (2\, \mathfrak{a} - \epsilon)}
\ee
which is the same result as that in \cite{IOYone} and coincides with the one instanton part of the instanton partition function.

Note that
\be
f_{N_L}(z) = \frac{1}{z^2} \sqrt{\beta} N_L ( \sqrt{\beta} N_L + \alpha_{1+2} - Q_E)
= \frac{1}{z^2} \frac{(\mathfrak{a}-m_2)(\mathfrak{a}+m_2)}{g_s^2},
\ee
\be
f_{N_R}(z) = - \frac{1}{z} \sqrt{\beta} N_R = \frac{1}{z} \frac{(\mathfrak{a}+m_1)}{g_s}.
\ee



\begin{thebibliography}{99}







\bibitem{IOYanok1}
  H.~Itoyama, T.~Oota and K.~Yano,
  ``Discrete Painlev\'{e} system and the double scaling limit of the matrix model for irregular conformal block and gauge theory,''
  Phys.\ Lett.\ B {\bf 789}, 605-609 (2019)
  [arXiv:1805.05057 [hep-th]].
    
\bibitem{Gai} 
	D.~Gaiotto,
	``Asymptotically free $\mathcal{N} = 2$ theories and irregular conformal blocks,''
	J.\ Phys.\ Conf.\ Ser.\  {\bf 462}, no. 1, 012014 (2013)
	[arXiv:0908.0307 [hep-th]].
	
	
\bibitem{MMM09092}
A.~Marshakov, A.~Mironov and A.~Morozov,
``On non-conformal limit of the AGT relations,''
Phys.\ Lett.\ B {\bf 682}, 125-129 (2009)
[arXiv:0909.2052 [hep-th]].


\bibitem{GT} 
D.~Gaiotto and J.~Teschner,
``Irregular singularities in Liouville theory and Argyres-Douglas type gauge theories, I,''
JHEP {\bf 1212}, 050 (2012)
[arXiv:1203.1052 [hep-th]].


	\bibitem{IOYone}
	H.~Itoyama, T.~Oota and N.~Yonezawa,
	``{Massive Scaling Limit of $\beta$-Deformed Matrix Model of Selberg Type},"
	Phys. Rev. D {\bf 82}, 085031 (2010) 
	[arXiv:1008.1861 [hep-th]].
	
	
\bibitem{DF} 
	V.~S.~Dotsenko and V.~A.~Fateev,
	``Conformal Algebra and Multipoint Correlation Functions in Two-Dimensional Statistical Models,''
	Nucl.\ Phys.\ B {\bf 240}, 312-348 (1984).
	
	
	
\bibitem{MMS10} 
	A.~Mironov, A.~Morozov and Sh.~Shakirov,
	``Conformal blocks as Dotsenko-Fateev Integral Discriminants,''
	Int.\ J.\ Mod.\ Phys.\ A {\bf 25}, 3173-3207 (2010)
	[arXiv:1001.0563 [hep-th]].
	
	
		\bibitem{IO5}
	H.~Itoyama and T.~Oota,
	``{Method of Generating $q$-Expansion Coefficients for Conformal Block and $\mathcal{N}=2$ Nekrasov Function by $\beta$-Deformed Matrix Model},"
	Nucl. Phys. B {\bf 838}, 298-330 (2010)
	[arXiv:1003.2929 [hep-th]].
	

	
\bibitem{BK} 
	E.~Br\'{e}zin and V.~A.~Kazakov,
	``Exactly Solvable Field Theories of Closed Strings,''
	Phys.\ Lett.\ B {\bf 236}, 144-150 (1990).
	
	
\bibitem{DS} 
	M.~R.~Douglas and S.~H.~Shenker,
	``Strings in Less Than One-Dimension,''
	Nucl.\ Phys.\ B {\bf 335}, 635-654 (1990).
	
	
\bibitem{GM} 
	D.~J.~Gross and A.~A.~Migdal,
	``Nonperturbative Two-Dimensional Quantum Gravity,''
	Phys.\ Rev.\ Lett.\  {\bf 64}, 127-130 (1990).
	
	
		\bibitem{AGT}
	L.~F.~Alday, D.~Gaiotto and Y.~Tachikawa,
	``{Liouville Correlation Functions from Four-dimensional Gauge Theories},"
	Lett. Math. Phys. {\bf 91}, 167-197 (2010)
	[arXiv:0906.3219 [hep-th]].
	
	
\bibitem{AD}
	P.~C.~Argyres and M.~R.~Douglas,
	``{New Phenomena in $SU(3)$ Supersymmetric Gauge Theory},"
	Nucl. Phys. B {\bf 448}, 93-126 (1995)
	[arXiv:hep-th/9505062].
	
	
\bibitem{APSW}
	P.~C.~Argyres, M.~R.~Plesser, N.~Seiberg and E.~Witten,
	``{New $\mathcal{N}=2$ Superconformal Field Theories in Four Dimensions},"
	Nucl. Phys. B {\bf 461}, 71-84 (1996)
	[arXiv:hep-th/9511154].
	
	
	
	
	
\bibitem{KY}
	T.~Kubota and N.~Yokoi,
	``{Renormalization Group Flow near the Superconformal Points in $\mathcal{N}=2$ Supersymmetric Gauge Theories},"
	Prog. Theor. Phys. {\bf 100}, 423-436 (1998)
	[arXiv:hep-th/9712054].
		
	
\bibitem{SW}
	N.~Seiberg and E.~Witten,
	``{Monopoles, Duality and Chiral Symmetry Breaking in $\mathcal{N}=2$ Supersymmetric QCD},"
	Nucl. Phys. B {\bf 431}, 484-550 (1994)
	[arXiv:hep-th/9408099].
	
\bibitem{HO} 
	A.~Hanany and Y.~Oz,
	``On the quantum moduli space of vacua of $N=2$ supersymmetric SU$(N_c)$ gauge theories,''
	Nucl.\ Phys.\ B {\bf 452}, 283-312 (1995)
	[hep-th/9505075].
	
	
\bibitem{GW80}
  D.~J.~Gross and E.~Witten,
  ``Possible third-order phase transition in the large-$N$ lattice gauge theory,''
  Phys.\ Rev.\ D {\bf 21}, 446-453 (1980).

\bibitem{wad1212} 
  S.~R.~Wadia,
  ``A Study of $U(N)$ Lattice Gauge Theory in 2-dimensions,''
  U. Cicago preprint EFI 79/44, July 1979 
  [arXiv:1212.2906 [hep-th]].

\bibitem{wad80}
  S.~R.~Wadia,
  ``$N = \infty$ phase transition in a class of exactly soluble model lattice gauge theories,''
  Phys.\ Lett.\  {\bf 93B}, 403-410 (1980).
  
 \bibitem{bes79} 
 D.~Bessis,
 ``A New Method in the Combinatorics of the Topological Expansion,''
 Commun.\ Math.\ Phys.\  {\bf 69}, 147-163 (1979).
 
 \bibitem{IZ80} 
 C.~Itzykson and J.~B.~Zuber,
 ``The planar approximation. II,''
 J.\ Math.\ Phys.\  {\bf 21}, 411-421 (1980).

\bibitem{PS90a}
	V.~Periwal and D.~Shevitz,
	``{Unitary-Matrix Models as a Exactly Solvable String Theories},''
	Phys. Rev. Lett. {\bf 64}, 1326-1329 (1990).
	
\bibitem{PS90b}
	V.~Periwal and D.~Shevitz,
	``{Exactly Solvable Unitary Matrix Models: Multicritical Potentials and Correlations},''
	Nucl. Phys. {\bf B344}, 731-746 (1990).
	

	
	
		\bibitem{Nek}
	N.~A.~Nekrasov,
	``{Seiberg-Witten Prepotential From Instanton Counting},"
	Adv. Theor. Math. Phys. {\bf 7}, 831-864 (2004)
	[arXiv:hep-th/0206161].
	
\bibitem{NY} 
	H.~Nakajima and K.~Yoshioka,
	``Instanton counting on blowup. 1.,''
	Invent.\ Math.\  {\bf 162}, 313-355 (2005)
	[arXiv:math/0306198 [math.AG]].
	
	
	\bibitem{GIL12} 
	O.~Gamayun, N.~Iorgov and O.~Lisovyy,
	``Conformal field theory of Painlev\'e VI,''
	JHEP {\bf 1210}, 038 (2012)
	[arXiv:1207.0787 [hep-th]].
	
\bibitem{GIL13} 
O.~Gamayun, N.~Iorgov and O.~Lisovyy,
``How instanton combinatorics solves Painlev\'{e} VI, V and IIIs,''
J.\ Phys.\ A {\bf 46}, 335203 (2013)
[arXiv:1302.1832 [hep-th]].
	
	
\bibitem{ILT} 
	N.~Iorgov, O.~Lisovyy and Y.~Tykhyy,
	``Painlev\'{e} VI connection problem and monodromy of $c=1$ conformal blocks,''
	JHEP {\bf 1312}, 029 (2013)
	[arXiv:1308.4092 [hep-th]].
	
\bibitem{nag1611} 
H.~Nagoya,
``Conformal blocks and Painlev\'{e} functions,''
arXiv:1611.08971 [math-ph].	
	
\bibitem{BLMST}
	G.~Bonelli, O.~Lisovyy, K.~Maruyoshi, A.~Sciarappa and A.~Tanzini,
	``On Painlev\'e/gauge theory correspondence,''
	Lett.\ Math.\ Phys.\ {\bf 107}, 2359-2413 (2017)
	[arXiv:1612.06235 [hep-th]].
	
	

	
	
\bibitem{GG} 
	A.~Grassi and J.~Gu,
	``Argyres-Douglas theories, Painlev\'{e} II and quantum mechanics,''
	JHEP {\bf 1902}, 060 (2019) 
	[arXiv:1803.02320 [hep-th]].
		


\bibitem{LNR1806} 
  O.~Lisovyy, H.~Nagoya and J.~Roussillon,
  ``Irregular conformal blocks and connection formulae for Painlev\'{e} V functions,''
  J.\ Math.\ Phys.\  {\bf 59}, no. 9, 091409 (2018)
  [arXiv:1806.08344 [math-ph]].

\bibitem{AJJRT1607} 
  S.~K.~Ashok, D.~P.~Jatkar, R.~R.~John, M.~Raman and J.~Troost,
  ``Exact WKB analysis of $ \mathcal{N} $ = 2 gauge theories,''
  JHEP {\bf 1607}, 115 (2016)
  [arXiv:1604.05520 [hep-th]].
  
\bibitem{MM} 
	A.~Mironov and A.~Morozov,
	``On determinant representation and integrability of Nekrasov functions,''
	Phys.\ Lett.\ B {\bf 773}, 34-46 (2017)
	[arXiv:1707.02443 [hep-th]].
	
	
\bibitem{MP90}
	R.~C. Myers and V.~Periwal,
	``{Exact Solution of Critical Self-Dual Unitary-Matrix Models},''
	Phys. Rev. Lett. {\bf 65}, 1088-1091 (1990).
	
\bibitem{ger54}
 Ya.~L.~Geronimus,
 \textit{Polynomials orthogonal on a circle and their applications},
 Amer.\ Math.\ Soc.\ Translation {\bf 1954} (1954), no. 104, 79 pp.;
 Translations, Ser. 1, Vol. 3, 
 Amer.\ Math.\ Soc.\ , 1-78 (1962) (reprinted).

\bibitem{ger60}
 Ya.~L.~Geronimus, 
 \textit{Polynomials Orthogonal on a Circle and Interval},
 Translated from the Russian by D.~E.~Brown, Ed. by I.~N.~Sneddon,
 International Series on Applied Mathematics, Vol. 18, 
 Pergamon Press, London (1960).

 
 
\bibitem{sim1}
 B.~Simon,
 \textit{Orthogonal Polynomials on the Unit Circle}, Part 1: Classical Theory,
 Colloquium Publications (Amer.\ Math.\ Soc.), Vol. 54,
 Amer.\ Math.\ Soc., Province, Rhode Island (2005).
 


\bibitem{IW0012}
 M.~E.~H.~Ismail and N.~S.~Witte,
 ``Discriminants and Functional Equations for Polynomials Orthogonal on the Unit Circle,''
 J.\ Approx.\ Theory {\bf 110}, 200-228 (2001)
 [arXiv:math/0012259 [math.CA]].
 
\bibitem{EM10}
	T.~Eguchi and K.~Maruyoshi,
	``{Seiberg-Witten theory, matrix model and AGT relation},"
	JHEP {\bf 1007}, 081 (2010)
	[arXiv:1006.0828 [hep-th]].
	
	
\bibitem{min9110} 
  J.~A.~Minahan,
  ``Flows and solitary waves in unitary matrix models with logarithmic potentials,''
  Nucl.\ Phys.\ B {\bf 378}, 501-522 (1992)
  [arXiv:hep-th/9111012].
  
\bibitem{his9611} 
  M.~Hisakado,
  ``Unitary matrix models with a topological term and discrete time Toda equation,''
  Phys.\ Lett.\ B {\bf 395}, 208-217 (1997)
  [arXiv:hep-th/9611177].

\bibitem{AMM1610}
  G.~\'{A}lvarez, L.~Mart\'{i}nez Alonso and E.~Medina,
  ``Complex saddle points in the Gross-Witten-Wadia matrix model,''
  Phys.\ Rev.\ D {\bf 94}, no. 10, 105010 (2016)
  [arXiv:1610.09948 [hep-th]].

\bibitem{oku1705}
  K.~Okuyama,
  ``Wilson loops in unitary matrix models at finite $N$,''
  JHEP {\bf 1707}, 030 (2017)
  [arXiv:1705.06542 [hep-th]].

\bibitem{his9609} 
  M.~Hisakado,
  ``Unitary matrix models and Painlev\'{e} III,''
  Mod.\ Phys.\ Lett.\ A {\bf 11}, 3001-3010 (1996)
  [arXiv:hep-th/9609214].

\bibitem{FW}
 P.~J.~Forrester and N.~S.~Witte,
 ``Discrete Painlev\'{e} equations, orthogonal polynomials on the unit circle, and $N$-recurrences
 for averages over $U(N)$ --- $\mathrm{P}_{\mathrm{III}{}'}$ and $\mathrm{P}_{\mathrm{V}}$ $\tau$-functions,''
 Int.\ Math.\ Res.\ Not.\ {\bf 2004}, Issue 4, 160-183 (2004)
 [arXiv:math-ph/0305029]. 

\bibitem{BH} 
E.~Br\'{e}zin and S.~Hikami,
``Duality and replicas for a unitary matrix model,''
JHEP {\bf 1007}, 067 (2010)
[arXiv:1005.4730 [hep-th]].


\bibitem{his9705} 
  M.~Hisakado,
  ``Unitary matrix models and phase transition,''
  Phys.\ Lett.\ B {\bf 416}, 179-183 (1998)
  [arXiv:hep-th/9705121].
     
\bibitem{min91a} 
  J.~A.~Minahan,
  ``Schwinger-Dyson equations for unitary matrix models with boundaries,''
  Phys.\ Lett.\ B {\bf 265}, 382-388 (1991).
  
\bibitem{min91b} 
  J.~A.~Minahan,
  ``Matrix models with boundary terms and the generalized Painlev\'{e} II equation,''
  Phys.\ Lett.\ B {\bf 268}, 29-34 (1991).

\bibitem{WMTB}
T.~T.~Wu, B.~M.~McCoy, C.~A.~Tracy and E.~Barouch,
``Spin-spin correlation functions for the two-dimensional Ising model: Exact theory in the scaling region,''
Phys.\ Rev.\ B {\bf 13}, 316-374 (1976). 


\bibitem{MPW}
B.~M.~McCoy, J.~H.~H.~Perk and T.~T.~Wu,
``Ising Field Theory: Quadratic Difference Equations for the $n$-Point Green's Functions on the Lattice,''
Phys.\ Rev.\ Lett.\  {\bf 46}, 757-760 (1981). 

\bibitem{Mc90}
B.~M.~McCoy,
``Spin systems, Statistical mechanics and Painlev\'{e} functions,''
in  \textit{Painlev\'{e} Transcendents}, D.~Levi and P.~Winternitz (eds),
NATO ASI Series B{\bf 278}, 377-391, Springer (1992).


\bibitem{FW02}
	P.~J.~Forrester and N.~S.~Witte,
	``Application of the $\tau$-function theory of Painlev\'{e} equations to random matries: 
	$\mathrm{P}_{\mathrm{V}}$, $\mathrm{P}_{\mathrm{III}}$, the LUE, JUE and CUE,''
	Commun.\ Pure Appl.\ Math.\ {\bf 55}, 679-727 (2002) 
	[arXiv:math-ph/0201051].

	
	
	
	
\bibitem{GRP91} 
  B.~Grammaticos, A.~Ramani and V.~Papageorgiou,
  ``Do Integrable Mappings Have the Painlev\'{e} Property?,''
  Phys.\ Rev.\ Lett.\  {\bf 67}, 1825-1828 (1991).
  
\bibitem{FGR93}
	A.~S.~Fokas, B.~Grammaticos and A.~Ramani,
	``From Continuous to Discrete Painlev\'{e} Equations,''
	J.\ Math.\ Anal.\ Appl.\ {\bf 180}, 342-360 (1993).
	
\bibitem{NSKGR96}
	F.~Nijhoff, J.~Satsuma, K.~Kajiwara, B.~Grammaticos and A.~Ramani,
	``A study of the alternate discrete Painlev\'{e} II equation,''
	Inverse Problems {\bf 12}, 697-716 (1996).

\bibitem{mar05}
  M.~Mari\~{n}o,
  \textit{Chern-Simons Theory, Matrix Models, and Topological Strings},
  Int.\ Ser.\ Monogr.\ Phys.\  {\bf 131}, Oxford University Press (2005).
  
\bibitem{gol80} 
  Y.~Y.~Goldschmidt,
  ``1/$N$ expansion in two-dimensional lattice gauge theory,''
  J.\ Math.\ Phys.\  {\bf 21}, 1842-1850 (1980).

\bibitem{mar0805} 
  M.~Mari\~{n}o,
  ``Nonperturbative effects and nonperturbative definitions in matrix models and topological strings,''
  JHEP {\bf 0812}, 114 (2008)
  [arXiv:0805.3033 [hep-th]].


\bibitem{LAG} 
L.~Alvarez-Gaum\'{e},
``Random surfaces, statistical mechanics and string theory,''
Helv.\ Phys.\ Acta {\bf 64}, 359-526 (1991).	

\bibitem{GM9304}
P.~H.~Ginsparg and G.~W.~Moore,
  ``Lectures on 2-D gravity and 2-D string theory,'' 
  in \textit{Recent Directions in Particle Theory: From Superstrings and Black Holes to Standard Model}, 
  Proceedings of TASI 1992, ed. by J. Harvey and J. Polchinski, 277-469, World Scientific (1993) 
  [arXiv:hep-th/9304011].
  

	
	
\bibitem{oka79}
  K.~Okamoto, 
  ``Sur les feuilletages associ\'{e}s aux \'{e}quation du second ordre \`{a} points critiques fixes de P. Painlev\'{e};
   Espaces des conditions initiales,''
  Japan J.\ Math. {\bf 5}, 1-79 (1979).
  
\bibitem{sak01}
  H.~Sakai,
  ``Rational Surfaces Associated with Affine Root Systems and Geometry of the Painlev\'{e} Equations,''
  Commun.\ Math.\ Phys.\ {\bf 220}, 165-229 (2001).
  
\bibitem{FN80} 
  H.~Flaschka and A.~C.~Newell,
  ``Monodromy- and Spectrum-Preserving Deformations I,''
  Commun.\ Math.\ Phys.\  {\bf 76}, 65-116 (1980).


\bibitem{mal22}
J.~Malmquist, 
``Sur les \'{e}quations diff\'{e}ntielles du second ordre dont l'int\'{e}gral g\'{e}n\'{e}ral a ses points critieue fixes,''
Ark.\ Math.\ Astr.\ Fys.\ {\bf 17}, 1-89 (1922-23).

\bibitem{oka86}
	K.~Okamoto,
	``Studies on the Painlev\'{e} Equations. III. Second and Fourth Painlev\'{e} Equations, 
	$\mathrm{P}_{\mathrm{II}}$ and $\mathrm{P}_{\mathrm{IV}}$,''
	Math.\ Ann. {\bf 275}, 221-255 (1986).

	
\bibitem{CKV0508}
 T.~Claeys, A.~B.~J.~Kuijlaars, and M.~Vanlessen,
 ``Multi-critical unitary random matrix ensembles and the general Painlev\'{e} II equation,''
  Ann.\ Math.\ {\bf 168}, 601-641 (2008) 
  [arXiv:math-ph/0508062].

\bibitem{GMN0907} 
  D.~Gaiotto, G.~W.~Moore and A.~Neitzke,
  ``Wall-crossing, Hitchin Systems, and the WKB Approximation,''
  Adv.\ Math.\ {\bf 234}, 239-403 (2013) 
  [arXiv:0907.3987 [hep-th]].

	
\bibitem{MMS91}
	M.~J. Bowick, A.~Morozov, and D.~Shevitz,
	``{Reduced unitary matrix models and the hierarchy of $\tau$-functions},''
	Nucl. Phys. {\bf B354}, 496-530 (1991).
	
\bibitem{miz0411} 
  S.~Mizoguchi,
  ``On unitary/Hermitian duality in matrix models,''
  Nucl.\ Phys.\ B {\bf 716}, 462-486 (2005)
  [arXiv:hep-th/0411049].

\bibitem{OV0205} 
  H.~Ooguri and C.~Vafa,
  ``World sheet derivation of a large $N$ duality,''
  Nucl.\ Phys.\ B {\bf 641}, 3-34 (2002)
  [hep-th/0205297].
 
 \bibitem{MMM09093}
 A.~Marshakov, A.~Mironov and A.~Morozov,
 ``Zamolodchikov asymptotic formula and instanton expansion in $\mathcal{N}=2$ SUSY $N_f=2N_c$ QCD,"
 JHEP {\bf 0911}, 048 (2009)
 [arXiv:0909.3338 [hep-th]].
  
	\bibitem{IMO} 
	H.~Itoyama, K.~Maruyoshi and T.~Oota,
	``The Quiver Matrix Model and 2d-4d Conformal Connection,''
	Prog.\ Theor.\ Phys.\  {\bf 123}, 957-987 (2010)
	[arXiv:0911.4244 [hep-th]].

\bibitem{EM09}
T.~Eguchi and K.~Maruyoshi,
``{Penner Type Matrix Model and Seiberg-Witten Theory},"
JHEP {\bf 1002}, 022 (2010)
[arXiv:0911.4797 [hep-th]].	
	
\bibitem{IYone}
	H.~Itoyama and N.~Yonezawa,
	``{$\epsilon$-Corrected Seiberg-Witten Prepotential Obtained From Half Genus Expansion in $\beta$-Deformed Matrix Model},"
	Int. J. Mod. Phys. A {\bf 26}, 3439-3467 (2011)
	[arXiv:1104.2738 [hep-th]].
	
\bibitem{NR1}
	T.~Nishinaka and C.~Rim,
	``{$\beta$-deformed matrix model and Nekrasov partition function},"
	JHEP {\bf 1202}, 114 (2012)
	[arXiv:1112.3545 [hep-th]].
	
\bibitem{NR2}
	T.~Nishinaka and C.~Rim,
	``{Matrix models for irregular conformal blocks and Argyres-Douglas theories},"
	JHEP {\bf 1210}, 138 (2012)
	[arXiv:1207.4480 [hep-th]].
	
\bibitem{CR}
	S.~K.~Choi and C.~Rim,
	``Parametric dependence of irregular conformal block,''
	JHEP {\bf 1404}, 106 (2014)
	[arXiv:1312.5535 [hep-th]].
	
\bibitem{MMS11} 
A.~Mironov, A.~Morozov and S.~Shakirov,
``Brezin-Gross-Witten model as 'pure gauge' limit of Selberg integrals,''
JHEP {\bf 1103}, 102 (2011)
[arXiv:1011.3481 [hep-th]].
	
\bibitem{Dai}
F.~David,
``Loop Equations and Nonperturbative Effects in Two-dimensional Quantum Gravity,''
Mod.\ Phys.\ Lett.\ A {\bf 5}, 1019-1030 (1990).

\bibitem{MM90} 
A.~Mironov and A.~Morozov,
``On the origin of Virasoro constraints in matrix models: Lagrangian approach,''
Phys.\ Lett.\ B {\bf 252}, 47-52 (1990).

	
\bibitem{IM}
	H.~Itoyama and Y.~Matsuo,
	``Noncritical Virasoro algebra of $d < 1$ matrix model and quantized string field,''
	Phys.\ Lett.\ B {\bf 255}, 202-208 (1991).	
	

	
	
		

	
	
\bibitem{IY} 
	H.~Itoyama and R.~Yoshioka,
	``Developments of theory of effective prepotential from extended Seiberg-Witten system and matrix models,''
	PTEP {\bf 2015}, no. 11, 11B103 (2015)
	[arXiv:1507.00260 [hep-th]].
	
	
\bibitem{Zam}
Al.~B.~Zamolodchikov,
``Conformal symmetry in two-dimensional space: Recursion representation of conformal block,"
Theor. Math. Phys. {\bf 73}, 1088-1093 (1987).	
	

\bibitem{Ake}
G.~Akemann,
``{Higher genus correlators for the hermitian matrix model with multiple cuts},"
Nucl.\ Phys.\  B {\bf 482}, 403-430 (1996)
[arXiv:hep-th/9606004].
	






\end{thebibliography}
\end{document}